\documentclass[11pt]{article}

\usepackage{PRIMEarxiv}

\usepackage[utf8]{inputenc} 
\usepackage[T1]{fontenc}    
\usepackage{hyperref}       
\usepackage{url}            
\usepackage{booktabs}       
\usepackage{amsfonts}       
\usepackage{nicefrac}       
\usepackage{microtype}      
\usepackage{lipsum}
\usepackage{fancyhdr}       
\usepackage{graphicx}       
\graphicspath{{media/}}     

\usepackage{lineno}
\usepackage{amsmath}
\usepackage{amssymb}
\usepackage{graphicx}
\usepackage{dcolumn}
\usepackage{bm}
\usepackage{siunitx}
\usepackage[normalem]{ulem}
\usepackage{float}

\newcommand{\mcj}{\mathcal{J}}

\newcommand{\tot}{\mathrm{tot}}
\newcommand{\zpf}{\mathrm{zpf}}
\newcommand{\eff}{\mathrm{eff}}
\newcommand{\myScale}{0.4}
\newcommand{\myScaleSmall}{0.27}

\title{Detection of a semitransparent object with no exchange of quanta}

\author{
  Vedran Vujnovi\'{c}\\
  Faculty of Physics and Center for Micro- and Nanosciences and Technologies \\
  University of Rijeka \\
 Radmile Matej\v{c}i\'{c} 2, 51000 Rijeka, Croatia\\
\And
  Nenad Kralj$^*$\\
  Faculty of Physics and Center for Micro- and Nanosciences and Technologies \\
  University of Rijeka \\
 Radmile Matej\v{c}i\'{c} 2, 51000 Rijeka, Croatia\\
$^*$\texttt{nenad.kralj@phy.uniri.hr} \\
\And
  Marin Karuza\\
  Faculty of Physics and Center for Micro- and Nanosciences and Technologies \\
  University of Rijeka \\
 Radmile Matej\v{c}i\'{c} 2, 51000 Rijeka, Croatia\\
}

\begin{document}
\maketitle

\begin{abstract}
In this paper, we theoretically analyze the optimization of a Fabry-P\'{e}rot cavity for the purpose of detecting partially absorbing objects placed inside without photon exchange. Utilizing the input-output formalism, we quantitatively relate the probability of correctly inferring the presence or absence of the object to the probability of avoiding absorption. We show that, if the cavity decay rate due to absorption by the object is comparable to that of the empty cavity and to the object-induced detuning, the product of the two probabilities is maximized by an undercoupled cavity, in which case detection in transmission is favorable to that in reflection. These results are contrary to the case of a perfect absorber, thus adding to the body of work pertaining to interaction-free measurement schemes and providing insight into optimizing their efficiency when detecting realistic objects.
\end{abstract}


\section{Introduction}
\label{sec:main1}
Various examples of interaction-free measurements were conceived to elucidate the measurement problem in quantum mechanics. The Renninger negative-result experiment~\cite{Renninger1953, Renninger1960} stipulates that a detection need not take place for a quantum measurement to occur. Elitzur and Vaidman~ (EV)~\cite{Elitzur_1993} refine the argument and provide a means of exploiting nonlocality. In an entangled pair, the state of one system can be inferred from afar by performing a measurement on its counterpart, assuming the experimenter is aware of the correlation. Conversely, EV propose to ascertain the presence of an object without interacting with it and with no prior knowledge about it, owing to the wavelike nature of a single quantum.
Their gedanken experiment uses a single-particle Mach-Zehnder interferometer to achieve this with \SI{50}{\percent} probability. Kwiat et al. have later shown how coherently repeated interrogation continually projects the system state and in principle enables a detection probability of \SI{100}{\percent}, as illustrated for a series of connected Mach-Zehnder interferometers and for two coupled Fabry-P\'{e}rot cavities~\cite{Kwiat_1995}. In particular, and in contrast to the EV proposal, a probability higher than \SI{50}{\percent} can be achieved even using coherent pulses with a mean photon number larger than unity. This was further studied by Pavi\v{c}i\'{c} for monolithic total-internal-reflection resonators~\cite{Pavicic:1996np, Paul:97, Paul1998RealisticID, Pavicic:2007}. In~\cite{Kwiat_1995}, the authors also show the first experimental realization of interaction-free detection, with a scheme which is however significantly simpler than that of the thought experiment, such that the obtained probabilities are still up to \SI{50}{\percent}. Instead, the first experiment featuring higher probabilities was done by Tsegaye et al.~\cite{FP_IFM_1998}, using strongly attenuated light from a laser diode in conjunction with a reasonably high-finesse Fabry-P\'{e}rot cavity. Taking into consideration also the detector inefficiencies and dark counts, they find at most an \SI{80}{\percent} probability of correctly ascertaining the object's presence inside the cavity or absence thereof, given the cavity parameters and provided the object was present with an \textit{a priori} probability of \SI{50}{\percent}.
It is important to note that some prior knowledge of the object is in reality still assumed by EV and in the subsequent works, in that it is taken to be perfectly absorbing or perfectly reflecting. Herein, we extend these ideas to realistic, semitransparent objects, using a Fabry-P\'{e}rot cavity as a detection platform. As argued in~\cite{Kwiat_1995}, a photon passing through an object generally acquires a phase and a subsequent detection of such a photon should, in a sense, not be considered interaction-free. For this reason, we restrict ourselves to interactions in which no quanta have been exchanged with the object, an interpretation of ``interaction-free'' employed earlier in~\cite{FP_IFM_1998}. As will be made apparent, optimizing the system for the detection of a semitransparent object is fundamentally different to a perfect absorber or reflector, since photons can decay into all possible detection ports even in its presence.  

\section{Input-output theory of optomechanical cavities}
\label{sec:main2}

\begin{figure}[h!]
    \centering
\includegraphics[width=0.63\textwidth]{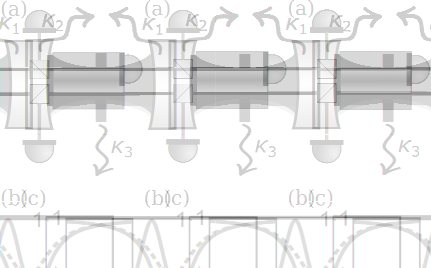}
 \vspace{-1mm}
    \caption{(a) Schematic of the proposed Fabry-P\'{e}rot setup for detecting a semitransparent object (black rectangle) with no exchange of quanta. The red beam is used for locking, while single photons or weak coherent pulses used for interaction-free detection are depicted in green. (b)~Transmission and (c)~reflection coefficients for a critically coupled empty cavity ($\xi = 0.5$, solid line) and an over- or under-coupled one ($\xi \neq 0.5$, dashed line). The mode matching efficiency is perfect ($\epsilon = 1$) in both cases. 
    }
    \label{fig:opticalCavityPortSchematic}
\end{figure}

To this end, we describe the system comprising an optical cavity and an object placed inside it, shown in Fig.~\ref{fig:opticalCavityPortSchematic}, within the paradigm of cavity optomechanics~\cite{aspelmeyer2014cavity}. Assuming that the optical mode interacts with a single mechanical mode, with their creation (annihilation) operators $a^\dagger (a)$ and $b^\dagger (b)$ respectively, the standard Hamiltonian describing dispersive optomechanical coupling can be written as $H_\mathrm{om} = \hbar g_0 a^\dagger a (b + b^\dagger)$. Here $g_0 = x_\zpf \cdot \mathrm{d}\omega_\mathrm{c}/\mathrm{d}x$  is the vacuum optomechanical coupling rate assuming ideal transverse overlap of the two mode shapes, $x$ is the object's position along the cavity axis, quantized as $x = x_\zpf (b + b^\dagger)$, with $x_\zpf$ the amplitude of mechanical zero-point fluctuations, and $\omega_\mathrm{c}$ is the cavity resonance frequency, now dependent on $x$. It can be shown that $g_0$ is maximal halfway between a node and an antinode of the intracavity intensity~\cite{Biancofiore2011}, at which point
$g_0^\mathrm{max} \approx 2 (\omega_\mathrm{c}/L) |r_\mathrm{m}| x_\mathrm{zpf}$, with $L$ the cavity length and $r_\mathrm{m}$ the membrane (field) reflectivity.
We are in fact not interested in studying the object's motion, but want to make use of the cavity resonance frequency shift due to its presence, which becomes apparent upon solving the Langevin equations of motion corresponding to the above Hamiltonian. With details of the derivation provided in Supplementary Information \ref{sec:SI1}, the steady-state amplitude of the intracavity field is found to be $\alpha = \sqrt{\kappa_1 \epsilon} a_\mathrm{in}/(\kappa/2 + i \Delta_P)$, where $a_\mathrm{in}$ is the input optical field amplitude, $\epsilon$ its modematching efficiency to the cavity mode and $\kappa = \kappa_1 + \kappa_2 + \kappa_3$ the total cavity (intensity) decay rate, with the three terms corresponding to decay through the input and opposing end-mirror, and that via absorption by the object, respectively (cf. Fig.~\ref{fig:opticalCavityPortSchematic}(a)). The former two loss rates are given by the mirrors' respective power transmissivities $T_j$ as $\kappa_j = (c/2L) T_j$, $j \in \{1,2\}$, with $c$ the speed of light in vacuum, while the latter loss
is naturally related to the object absorptivity but is dependent on its position with respect to the standing wave and calculated as the imaginary part of the cavity resonance frequency shift (see, e.g.,~\cite{Biancofiore2011}). Importantly, the detuning $\Delta_P = \Delta_A - 2 (g_0 |\alpha|)^2/\omega_\mathrm{m}$ now incorporates the optomechanical frequency shift compared to the empty cavity value $\Delta_A = \omega_\mathrm{c} - \omega_\mathrm{L}$ where $\omega_\mathrm{L}$ is the input laser frequency. This total detuning depends on the object reflectivity and position via $g_0$.

From the perspective of probing the object's presence using single photons or weak coherent pulses, what is of interest are the probabilities of a particular photon being reflected, transmitted or absorbed by the compound cavity-object system. In our framework, these are analogous to the corresponding cavity coefficients, which follow from the output amplitudes in the three decay ports, in turn given by the input-output relations~\cite{GardinerZoller, aspelmeyer2014cavity}, $a_{\mathrm{out},k} = a_{\mathrm{in},k} - \sqrt{\kappa_k} \alpha$, with $k \in \{1,2,3\}$, $a_{\mathrm{in},1} \equiv a_\mathrm{in}$ and $a_{\mathrm{in},2} = a_{\mathrm{in},3} = 0$, since we take the cavity to be pumped only from port~1 and consider the quantum and classical noise as negligible. Defining a given cavity coefficient as the ratio $P_k/P_\mathrm{in}$, with $P_k = |a_{\mathrm{out},k}|^2$ the output power in the relevant port and $P_\mathrm{in} = |a_\mathrm{in}|^2$ the input power, we have
\begin{align}
    \mathcal R = 1 - \frac{\epsilon \kappa_1 (\kappa - \kappa_1)}{(\kappa/2)^2 + \Delta^2} \, , \quad \mathcal T = \frac{\epsilon \kappa_1 \kappa_2}{(\kappa/2)^2 + \Delta^2} \, , \quad \mathcal A = \frac{\epsilon \kappa_1 \kappa_3}{(\kappa/2)^2 + \Delta^2} \, ,
    \label{eq:cavCoeff}
\end{align}
for reflection, transmission and absorption, i.e. ports 1, 2 and 3 respectively. See Supplementary Information \ref{sec:SI1} for details. We have implicitly assumed
the inaccessible loss is exclusively due to object absorption, as opposed to
mirror absorption and scattering.
Also note
that $\Delta$, $\kappa$, $\epsilon$, and by extension the cavity coefficients, depend on the object's presence or absence. Here they are all ambivalent in that sense but, where necessary to make a distinction, we will use the subscripts ``$P$'' and ``$A$'' to indicate the two respective states. In state $A$, $\kappa_3 = g_0 = 0$, entailing $\kappa_A = \kappa_1 + \kappa_2$, $\mathcal A \rightarrow 0$ and $\Delta \rightarrow \Delta_A$ as defined before. Finally, for the following discussion we will express the mirror loss rates in terms of the empty cavity coupling efficiency $\xi = \kappa_1/(\kappa_1 + \kappa_2)$ as $\kappa_1(\xi) = \xi \kappa_A$ and $\kappa_2(\xi) = (1-\xi)\kappa_A$.

\section{Figures of merit for interaction-free detection of a semitransparent object}
\label{sec:main3}

The first figure of merit for interaction-free detection via the described system we have termed the \emph{security}, namely the probability of avoiding absorption when the object is there. For a single photon, it is $\eta = (\mathcal R_P + \mathcal T_P)/(\mathcal R_P + \mathcal T_P + \mathcal A) = 1 - \mathcal A$, where we have used
$\mathcal R_P + \mathcal T_P + \mathcal A = 1$.
Now, if the object is very absorptive, i.e. $\kappa_3 \gg \kappa_1,\kappa_2$,
$\mathcal T_P \rightarrow 0$
and $\eta$ can only be improved by reducing the transmissivity of the input mirror. Doing so without making the other mirror correspondingly less transmissive, thus creating an asymmetrical cavity ($\xi \neq 0.5$), amounts to diminishing the discriminant between the object's presence and absence, whether the detection is done in transmission or in reflection (cf. Fig.~\ref{fig:opticalCavityPortSchematic}(b) and (c)).
In other words, the functional dependence $\eta(\xi)$ is largely immaterial if the object is a perfect black body. Assuming the cavity is critically coupled ($\xi=0.5$) and resonant with no object inside, a signal photon can then only be detected in reflection if the object is present and in transmission otherwise. Given a certain a priori probability of the object being present, as well as the detector efficiencies and dark counts, one can only evaluate the probability of detecting this photon, as detailed in~\cite{FP_IFM_1998}.

The situation is fundamentally different for a semitransparent object. Such an object need not absorb a photon transmitted through the input coupler and it therefore might be advantageous to increase the chance of the photon decaying through the transmission port, i.e. introduce an asymmetry. This would
still come at the expense of the probability of correctly ascertaining the object's presence,
since a given photon would have a finite probability of being transmitted as well as reflected by the cavity in either case, $A$ or $P$. In order to study this systematically, we take as our signal the difference between the photon counts in the
two cases
for either output, and correspondingly define the \emph{signal-to-noise ratio}~(SNR) as
\begin{align}
    \mathrm{SNR}_j(\xi,N_0) = \frac{\sqrt{N_0} \chi_j |\mcj_A(\xi) - \mcj_P(\xi)|}{\sqrt{ \chi_j [\mcj_A(\xi) + \mcj_P(\xi)] + 2 D_j}} \, ,
    \label{eq:SNR}
\end{align}
where $j \in \{1,2\}$ pertains to the two accessible output ports and $\mathcal J$ is the related cavity coefficient, $\mathcal J \in \{\mathcal R, \mathcal T\}$. 
The total number of input photons $N_0 = C_0 t$ can be expressed as a product of the flux $C_0$ and measurement time $t$, $\chi_j$ is the quantum efficiency of the detector in a given port and $D_j = C_j/C_0$
with $C_j$ its dark count rate. Supplementary Information \ref{sec:SI2} contains the derivation of the presented SNR form. In particular, we have assumed the photons obey Poissonian statistics, such that the uncertainty of each measurement scales with $\sqrt{N_0}$, and the SNR inherits this scaling.
From the SNR, one can infer the probability of 
estimating the object's presence based on a particular data run. The more photons impinge on the cavity, the less likely it is that none of them will be absorbed, which can be quantified by the total security
\begin{align}
    \eta_\tot(\xi, N_0) = [1 - \mathcal{A}(\xi)]^{N_0} \, .
    \label{eq:totalSecurity}
\end{align}
It follows from bEq.~(\ref{eq:cavCoeff}) that both the SNR and $\eta_\tot$ depend on the amount of detuning and mode-mismatch resulting from the introduction of the object inside the cavity. Henceforth, we will assume that the incoming photons are perfectly matched to, and locked on resonance with, the empty cavity mode, i.e. $\epsilon_A =1$, $\Delta_A=0$. In practice, $\Delta_A = 0$ can be ensured by using cavity end-mirrors with coatings highly reflective for two different wavelengths, with the beam at one wavelength locked to the cavity. The cavity lock is propagated to the photons at the other wavelength by independently locking them to this beam. Stipulating the object's reflectivity is negligible at the cavity-locking wavelength as compared to that of the single photons, $\Delta_P \rightarrow \Delta_A$ for the former. Maximizing $g_0$ by placing the object between a node and an antinode of the intracavity standing wave, $\Delta_P$ for the single photons can be made significantly different from $\Delta_A$.
Regarding mode matching, $\eta_\tot$ and SNR in either port are both maximized for $\epsilon_P \rightarrow 0$. In that case, $\eta_\tot \rightarrow 1$ and the SNR can be made arbitrarily large by increasing $N_0$. In our analysis, we instead assume a finite $\epsilon_P$ and, for a given set of $\kappa_A$, $\kappa_3$ and $\Delta_P$, study the dependence of the system performance as an interaction-free detector on $\xi$ and $N_0$. Exploiting the results in practical applications requires precise a priori knowledge of the object the presence of which is to be ascertained. This might seem contrary to the idea of EV, but they too assume knowledge of the object's absorptivity.
Furthermore, we have derived the cavity coefficients using steady-state solutions for $\alpha$ and $x$.
The validity of the analysis is contingent on ensuring that on average at most one photon is resonating inside the cavity at a given time, thus realizing a quasi-steady state. This implies an upper bound on the allowed photon flux as $C_0^\mathrm{max} = \kappa_P/(2\pi)$. We parameterize the SNRs in terms of $N_0$ and $D_j$, while $C_0$ does not enter explicitly. However, the dark count ratios are fixed to $D_1 = D_2 = \num{e-3}$ throughout, which for typical detectors corresponds to $C_0 \sim \SI{e6}{\per\second}$, meeting the criterion for all $\kappa_P$ considered.

\section{Results}
\label{sec:main4}

\begin{figure}[h!]
    \centering
    \includegraphics[width=0.63\textwidth]{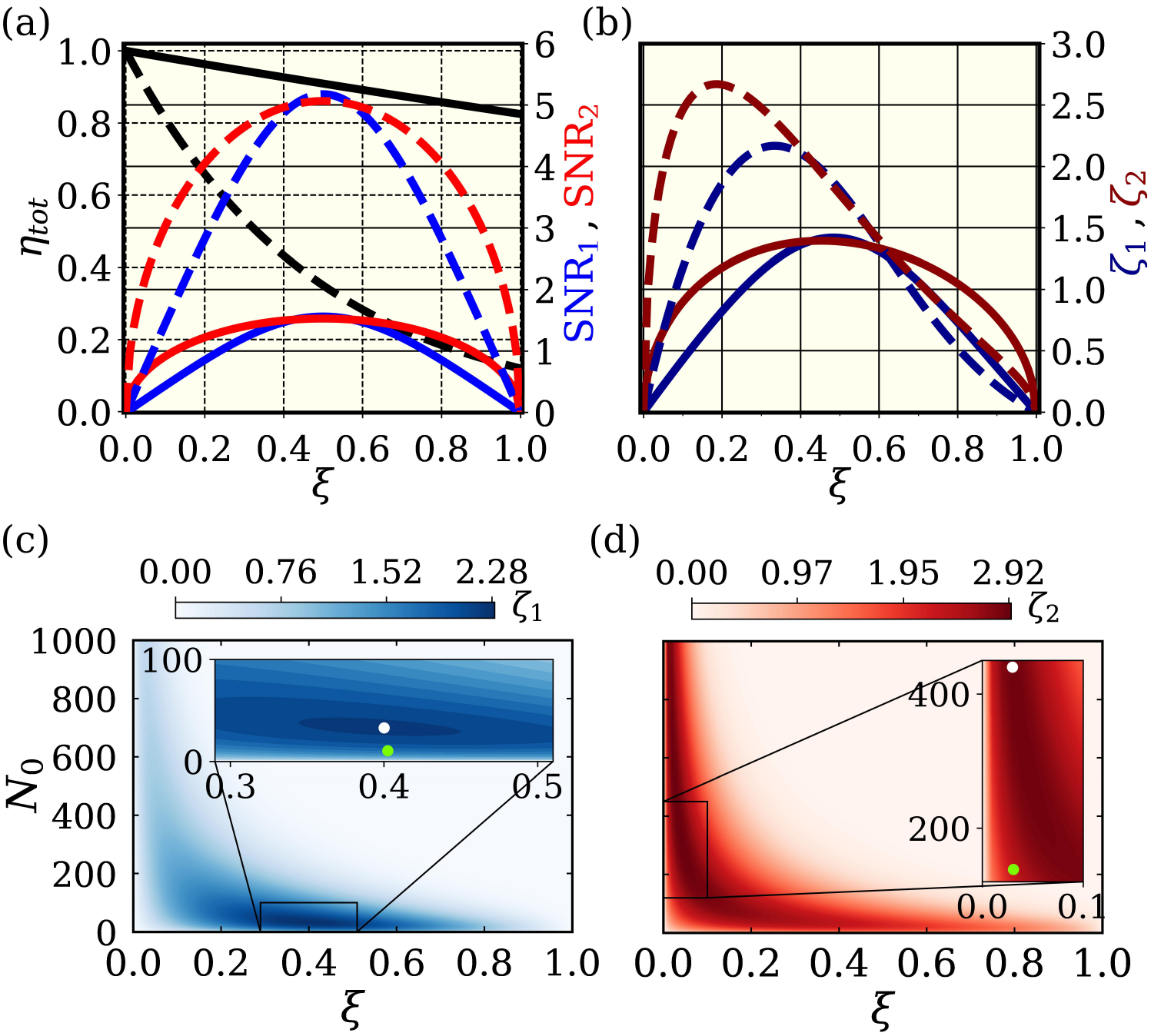}
     \vspace{-1mm}
    \caption{(a)~
    Total security (black) and SNRs in cavity reflection (blue) and transmission (red).~(b) The products $\zeta_1$ (dark blue) and $\zeta_2$ (dark red). In both panels solid lines correspond to $N_0 = 5$ and dashed lines to $N_0 = 55$ photons. The other cavity and object parameters are
    $\kappa_A/(2\pi) = \SI{1.5e7}{\hertz}$,  $\kappa_3/(2\pi) = \SI{6.5e6}{\hertz}$, $\Delta_P/(2\pi) = \SI{2e7}{\hertz}$, $\epsilon_P = 0.2$, $\chi_{1} = \chi_{2} = 0.5$ and $D_1 = D_2 = \num{e-3}$. (c)~Products $\zeta_1$ and (d)~$\zeta_2$, over a region of the $\xi N_0$-space, with the other system parameters set as in~(a) and (b). The insets show the global maxima and conditional maxima with $\eta_\tot \geq 0.85$ and $\mathrm{SNR}_j \geq 2$, denoted as white and green circles, respectively.
    }
    \label{fig:etaS_SNRs_independed-2Cases}
\end{figure}

Fig.~\ref{fig:etaS_SNRs_independed-2Cases}(a) shows $\eta_\tot$ and SNR$_j$ as functions of $\xi$, for two distinct values of $N_0$ and other parameters fixed as specified in the caption. Particularly, in this case the cavity is approximately as lossy via absorption as it is via the accessible ports, i.e. $\kappa_3 \approx \kappa_A$, and also $\kappa_3 \approx \Delta_P$. There is a twofold trade-off between security and probability of correct inference of the object's presence.
On the one hand, SNRs are maximal for $\xi = 0.5$, while $\eta_\tot$ is maximal for a perfectly reflective input mirror. On the other hand, SNRs are improved the more the cavity is probed, i.e. with increasing $N_0$, whereas $\eta_\tot$ falls off faster with increasing $\xi$ for a larger $N_0$. We choose to take the SNR and $\eta_\tot$ on equal footing and investigate their product $\zeta_j(\xi,N_0)$. 
Fig.~\ref{fig:etaS_SNRs_independed-2Cases}(b) shows that there are such $N_0$ for which the product is maximized by an asymmetrical cavity. Moreover, for some $N_0$ the maximal product is larger in transmission, and for others in reflection. Both of these features are of particular interest, since they represent clear distinctions from the case of a perfect black body, and neither are artefacts of the detectors in reflection or transmission, as the two are taken to be the same.

\begin{figure}[h!]
    \centering
    \includegraphics[width=0.64\textwidth]{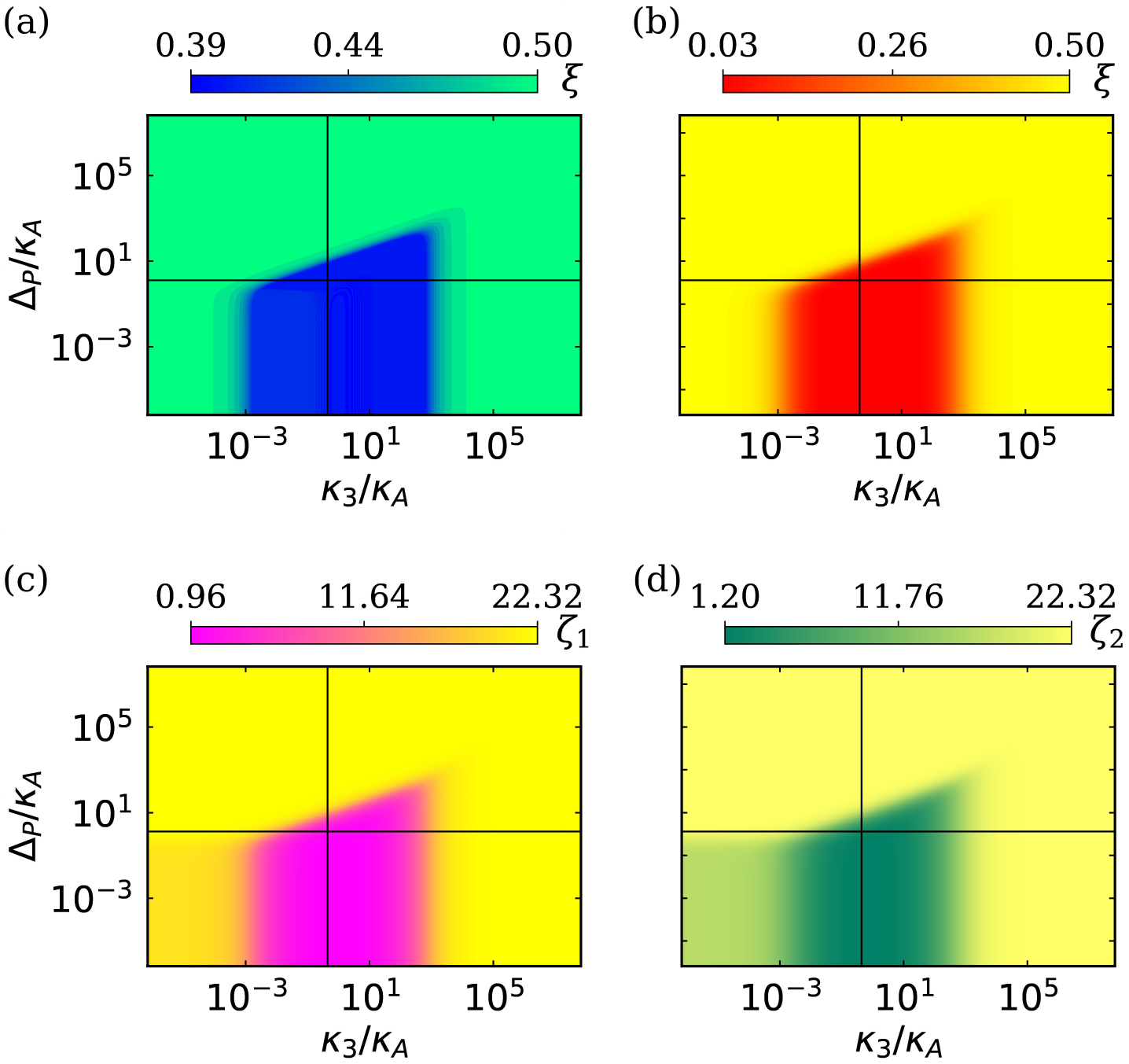}
     \vspace{-2mm}
    \caption{Cavity coupling efficiency that realizes the global maximum of (a)~$\zeta_1$ and (b)~$\zeta_2$ for a range of $\kappa_3$ and $\Delta_P$. The empty cavity decay rate is fixed to $\kappa_A/(2\pi) = \SI{1.5e7}{\hertz}$ and the mode matching to $\epsilon_P = 0.2$. (c)~Values of the maxima of $\zeta_1$ and (d)~$\zeta_2$ corresponding to panels (a) and (b). In all panels, 
    solid black lines indicate the $\kappa_3$ and $\Delta_P$ values corresponding to the system investigated in Figs.~\ref{fig:etaS_SNRs_independed-2Cases} and \ref{fig:etaTOT_as_a_function_of_SNRs_for_xOPT}.
    } 
    \label{fig:etaS_SNRs_product_surfacePlot}
\end{figure}

At this point one might still think that for a yet higher $N_0$ the maxima of $\zeta_j$ could skew back towards a critically coupled cavity. To exclude this possibility, we extend the search for the maxima to a region of the two-dimensional parameter space spanned by $\xi$ and $N_0$. For a fixed $\xi$, there is exactly one maximum in $N_0$. If it can be found for each $\xi$ within the swept subspace, and $\xi$ spans the entire range, the local maximum of the subspace is the global one. Instead of the absolute global maximum, one could search for a conditional maximum with constraints on the lowest acceptable $\eta_\tot$ and/or SNR. Figs.~\ref{fig:etaS_SNRs_independed-2Cases}(c) and \ref{fig:etaS_SNRs_independed-2Cases}(d) show the products for transmission and reflection, indicating in particular their global and conditional maxima with the constraints $\eta_\tot \geq 0.85$ and $\mathrm{SNR}_j \geq 2$.
System parameters are the same as in Fig.~\ref{fig:etaS_SNRs_independed-2Cases}(b), such that the latter corresponds to cross-sections of the former two at $N_0 \in \{5, 55\}$. The conditional maxima appear at $\xi < 0.5$ for both $\zeta_1$ and $\zeta_2$, with the latter being significantly more undercoupled ($\xi \approx \num{0.03}$ compared to $\xi \approx \num{0.4}$), and also featuring a somewhat higher maximum value. The global maxima do not fulfill the aforementioned constraints, but they are importantly also found within the investigated region, and shifted away from a critically coupled cavity in a similar manner.
Products analogous to those in Figs.~\ref{fig:etaS_SNRs_independed-2Cases}(c) and \ref{fig:etaS_SNRs_independed-2Cases}(d) for different ratios between $\kappa_A$, $\kappa_3$ and $\Delta_P$ are included in Supplementary Information \ref{sec:SI3}. Those results are summarized here in Fig.~\ref{fig:etaS_SNRs_product_surfacePlot}, panels (a) and (b) of which show the value of $\xi$ at which the global maxima of $\zeta_j$ appear as $\kappa_3$ and $\Delta_P$ are varied with respect to $\kappa_A$ for each cavity output. Clearly, in the majority of cases $\zeta_j$ are largest for a critically coupled cavity. However, there is a region in parameter space approximately centered around $\kappa_A \approx \kappa_3$ and with $\Delta_P \lesssim \kappa_A$ for which a (very) undercoupled cavity is in this sense preferable. In practice, one might be limited to this regime in the case of a weakly absorbing object for which $\kappa_A \ll \kappa_3$ is not readily accessible, or in the bad cavity limit with a very good absorber. Perhaps unsurprisingly, the values of the products $\zeta_j$, shown in Figs.~\ref{fig:etaS_SNRs_product_surfacePlot}(c) and (d) over the same parameter range, suggest that where the maxima occur for an undercoupled cavity, detection in transmission is slightly better than that in reflection. Still, the maxima are significantly diminished with respect to those in regions where a critically coupled cavity is best.

\begin{figure}[h!]
    \centering
    \includegraphics[width=0.6\textwidth]{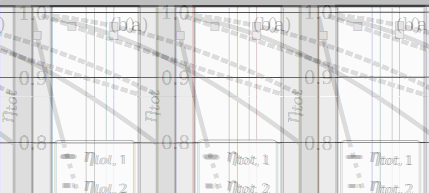}
    \vspace{-1mm}
    \caption{Total security as a function of (a)~SNR and (b)~number of impinging photons for detection done in reflection (solid blue line) and transmission (dashed red line), for the system investigated in Fig.~\ref{fig:etaS_SNRs_independed-2Cases}. The coupling efficiencies are set to values corresponding to maxima of $\zeta_1$ and $\zeta_2$ conditioned on $\eta_\tot \geq 0.85$ and $\mathrm{SNR}_j \geq 2$, which are herein indicated by yellow and grey squares respectively. The dotted blue line in 
    (b) is an extrapolation beyond $\mathrm{SNR = 5}$.
    }
    \label{fig:etaTOT_as_a_function_of_SNRs_for_xOPT}
\end{figure}

We now again turn to a particular system, i.e one with specified $\kappa_3$, $\Delta_P$ and $\kappa_A$. Once the (conditional or global) maxima of $\zeta_j$ have been found, we can set $\xi$ to the optimal value (independently for transmission and reflection) and investigate how $\eta_\tot$ scales with the SNR we wish to obtain. For this purpose, $N_0$
is expressed as a function of SNR from Eq.~(\ref{eq:SNR}) and substituted into Eq.~(\ref{eq:totalSecurity}). Fig.~\ref{fig:etaTOT_as_a_function_of_SNRs_for_xOPT}(a) shows the result for the object, cavity and detector parameters considered in Fig.~\ref{fig:etaS_SNRs_independed-2Cases} and $\xi$ fixed to the value pertaining to the conditional maxima found in Figs.~\ref{fig:etaS_SNRs_independed-2Cases}(c) and \ref{fig:etaS_SNRs_independed-2Cases}(d). This is clearly an example of a semitransparent object-cavity system for which detecting the object via transmission of the cavity outperforms doing it in reflection for any SNR or desired level of security, a prospectless situation with a perfect absorber. Fig.~\ref{fig:etaTOT_as_a_function_of_SNRs_for_xOPT}(b) instead expresses $\eta_\tot$ as a function of $N_0$, suggesting that the detection scheme featuring a larger maximum of $\zeta$ can be expected to take a longer time to reach it.
Further note that $\mathrm{SNR_j} \geq 1$ can be obtained even for $\eta_\tot \rightarrow 1$. In the other extreme, the monotonous roll-off suggests that, for finite $\Delta_P$ and $\epsilon_P$ investigated here, there are maximal $\mathrm{SNR}_j$ for interaction-free detection, which correspond to $\eta_\tot \rightarrow 0$. The roll-off can be made more favorable with increasing $\Delta_P$ and reducing $\epsilon_P$ and $\kappa_3$, and the maximal probability of inference can be approximated as unity for all practical applications, mimicking the idealized situation discussed in~\cite{Kwiat_1995}. 

\section{Conclusion}
\label{sec:main5}

Interaction-free measurements have been adopted in both fundamental research, e.g. quantum engines~\cite{Elouard2020} and the Hardy paradox~\cite{Hardy1992}, as well as practical applications like probing light-sensitive objects via interaction-free ghost imaging~\cite{Zhang2019}. Their potentiality is further highlighted by realizations having been demonstrated in the microwave domain~\cite{Dogra2022} and those not requiring single-particle sources~\cite{Peise2015,FP_IFM_1998}. Our study is aimed at aiding these efforts by maximizing the effectiveness of the pertinent schemes, focusing on a Fabry-P\'{e}rot cavity with a realistic, semitransparent object. The developed model relating the probability of correctly estimating the object's presence to the security of the process, is wavelength-agnostic, i.e. holds for both optical and microwave domains. With the model, we find that, unlike for a perfect black body, an undercoupled cavity might be advantageous for detecting a semitransparent object, depending on how absorptive it is relative to the cavity mirrors' transmissivities. We further demonstrate that in such a case cavity transmission leads to more efficient detection than reflection.
These results can be experimentally tested using a thin SiN membrane placed inside a cavity with end-mirrors featuring similarly high reflectivities at \SI{532}{\nano\meter} and \SI{1064}{\nano\meter}, in conjunction with a laser with outputs at both wavelengths. The beam at \SI{1064}{\nano\meter} can be locked on resonance to the cavity, and the intermediate lock between it and the single photons at \SI{532}{\nano\meter} is ensured by the laser's built-in frequency doubling crystal. The membrane can be metallized so as to exhibit near-zero reflectivity and absorptivity at the locking wavelength and moderate ones at the single photon wavelength. Specifically, the latter absorptivity could be made sufficiently large for the entire system to be operable in the interesting regime where undercoupled cavities are preferable.

From a theoretical standpoint, we finally note that we have only considered SNRs for single detectors placed either in transmission or in reflection, derived by comparing their clicks when the object is absent versus when it is present. In fact, a more complete treatment could be made to include both detectors simultaneously, leading to an improvement of the system's performance regardless of how the object parameters compare to those of the employed cavity.

\section*{Acknowledgments}
This study was supported by the University of Rijeka grant no. 23-38, Croatian Science Foundation grant IP-2014-09-7515, Ministry of Science and Education (MSE) of Croatia Contract No. KK.01.1.1.01.0001.
Simulations were performed using the ``Bura'' supercomputer of the University of Rijeka Center for Advanced Computing and Modelling.

\bibliographystyle{unsrt}  
\bibliography{templateArxiv}

\newpage 

\textbf{\hspace{5.0cm}\Large{Supplementary Information}}

\setcounter{section}{0} 
\renewcommand{\thesection}{\Alph{section}} 

\section{Cavity coefficients in the input-output formalism}
\label{sec:SI1}

We will describe the Fabry-P\'{e}rot cavity with an object placed inside (cf. Fig.~\ref{fig:opticalCavityPortSchematic}(a) of the main text) within the paradigm of optomechanics~\cite{aspelmeyer2014cavity}. Specifically, the model provided here encapsulates all relevant effects for a canonical system of a cavity with a moving end-mirror. Strictly speaking, our system is a ``membrane-in-the-middle''~\cite{Thompson2008}, for which the dependence of the resonance frequency on the membrane (object) position is found from the transfer matrix model assuming the fields are plane waves~\cite{Jayich_2008} (see~\cite{Dumont2019} for the case of a membrane at the edge), or in an even more rigorous way, as in~\cite{Cheung2011, Biancofiore2011}. However, the canonical model is sufficient to argue when asymmetrical cavities are beneficial in the context of interaction-free detection, and we thus employ it here. 

If the object is mechanically compliant, the full Hamiltonian can be written as~\cite{aspelmeyer2014cavity}
 \begin{align}
  H_\mathrm{tot} = \hbar \omega_\mathrm{c}(x)a^\dagger a + \hbar \omega_\mathrm{m} b^\dagger b + \hbar \sqrt{\kappa_1} a_\mathrm{in}(a e^{i\omega_\mathrm{L} t} + a^\dagger e^{-i\omega_\mathrm{L} t}) \, ,
 \label{eq:ham}
 \end{align}
where we have assumed that the optical cavity normal mode interacts with a single mechanical mode, and their creation (annihilation) operators are $a^\dagger (a)$ and $b^\dagger (b)$ respectively. The reduced Planck constant is denoted by $\hbar$, $\omega_\mathrm{c}$ and $\omega_\mathrm{m}$ are the frequencies of the two respective modes, $a_\mathrm{in}$ is the amplitude of the input pump laser, $\omega_\mathrm{L}$ its frequency and $\kappa_1$ the optical (intensity) decay rate through the input mirror.
The underlying interaction is that of radiation pressure, which depends on the object's position $x$ with respect to the intracavity standing wave, as evident from the optical mode frequency being a function of $x$.
The Taylor expansion of this function up to the term linear in $x$ yields
 \begin{align}
  H_\mathrm{tot} = \hbar \omega_\mathrm{c}(0)a^\dagger a + \hbar \omega_\mathrm{m} b^\dagger b + \hbar G x a^\dagger a + \hbar \sqrt{\kappa_1} a_\mathrm{in}(a e^{i\omega_\mathrm{L} t} + a^\dagger e^{-i\omega_\mathrm{L} t}) \, ,
 \end{align}
with $\omega_\mathrm{c}(0)$ the empty cavity resonance frequency and $G = \text{d}\omega_\mathrm{c}/\text{d}x$ the so-called frequency pull parameter. It is convenient to move into the reference frame rotating at the driving laser frequency, amounting to
\begin{align}
  H_\mathrm{tot} = \hbar \Delta_A a^\dagger a + \hbar \omega_\mathrm{m} b^\dagger b + \hbar G x a^\dagger a + \hbar \sqrt{\kappa_1} a_\mathrm{in}(a + a^\dagger) \, ,
 \end{align}
where $\Delta_A = \omega_\mathrm{c}(0) - \omega_\mathrm{L}$.
Further quantizing the mechanical displacement according to $x = x_{\zpf}(b + b^\dagger)$ with $x_{\zpf} = \sqrt{\hbar/(2 m_{\eff} \omega_\mathrm{m})}$ the amplitude of the displacement zero-point fluctuations and $m_{\eff}$ the object effective mass, and defining the vacuum optomechanical coupling rate as $g_0 = G x_{\zpf}$ finally leads to the form
\begin{align}
  H_\mathrm{tot} = \hbar \Delta_A a^\dagger a + \hbar \omega_\mathrm{m} b^\dagger b + \hbar g_0 a^\dagger a (b + b^\dagger) + \hbar \sqrt{\kappa_1} a_\mathrm{in}(a + a^\dagger) \, .
  \label{eq:Hfinal}
 \end{align}
In the coupling rate $g_0$, we have also assumed a perfect transverse mode overlap between the optical and mechanical mode. It can be shown~\cite{Biancofiore2011} that $g_0$ is maximal halfway between a node and an antinode of the intracavity intensity, at which point
$g_0^\mathrm{max} \approx 2 (\omega_\mathrm{c}/L) |r_\mathrm{m}| x_\mathrm{zpf}$, with $L$ the cavity length and $r_\mathrm{m}$ the membrane (field) reflectivity. The implicit assumption there is that the object thickness is small compared to the period of the intracavity intensity $\lambda_\mathrm{c}/2$, with $\lambda_\mathrm{c}$ the resonant field wavelength. 

Employing the Heisenberg equation for the time evolution of operators with the Hamiltonian from Eq.~(\ref{eq:Hfinal}), and including dissipation and noise terms, leads to the Langevin equations of motion for $a$ and $b$. We will treat the (classical and quantum) noise terms and negligible, with which it follows that
\begin{align}
    \dot a &= -\left[ \kappa/2 + i\Delta_A + i g_0(b + b^\dagger) \right] a + \sqrt{\kappa_1}a_\mathrm{in} \, ,  \label{eq:eom} \\
    \dot b &= - i \omega_\mathrm{m} b - i g_0 a^\dagger a \, ,     \label{eq:eom2}
\end{align}
where $\kappa = \kappa_1 + \kappa_2 + \kappa_3$ is the total cavity decay rate, with $\kappa_2$ pertaining to the second mirror and $\kappa_3$ to absorption by the object. These EOMs are typically linearized around the steady-state values, with the dynamics retained in the small fluctuations about them. For describing the system as an interaction-free detector, we are not interested in studying the fluctuations, but simply in the steady-state values, $\alpha$ and $\beta$, obtained by setting $\dot a = \dot b = 0$. We find
\begin{align}
     \alpha &= \frac{\sqrt{\kappa_1} a_\mathrm{in}^\parallel}{\kappa/2 + i \Delta} \, , \\
    \beta &= - \frac{g_0 |\alpha|^2}{\omega_\mathrm{m}} \, ,
\end{align}
where $a_\mathrm{in}^\parallel = \sqrt{\epsilon} a_\mathrm{in}$ accounts for the fact that the input field is in general mode-matched to the cavity mode with some efficiency $\epsilon$, i.e. $a_\mathrm{in} = \sqrt{\epsilon} a_\mathrm{in}^\parallel + \sqrt{1 - \epsilon} a_\mathrm{in}^\bot$, with $a_\mathrm{in}^\bot$ the non-mode-matched part. The detuning $\Delta = \Delta_A + 2 g_0 \beta = \Delta_A - 2 g_0^2 |\alpha|^2/\omega_\mathrm{m}$ has been shifted due to the optomechanical interaction, i.e. due to the radiation pressure pushing the object and establishing a new equilibrium position. This shift depends on the position and reflectivity via $g_0$. If $g_0 \neq 0$, the presence of the object is implied, as indicated by the subscript ``$P$'' (and ``$A$'' standing for its absence), such that $\Delta \rightarrow \Delta_P$. However, the expressions for the steady-state amplitudes and, by extension, the cavity coefficients derived below, are valid for both states, $A$ and $P$, for which reason we omit the subscript in $\Delta$ here. The same goes for $\kappa$, as $\kappa_3 = 0$ with no object, and $\epsilon$, as its introduction in the cavity would generally change the mode matching.

With this, we can derive the optomechanical cavity reflection, transmission and absorption coefficients. Firstly, the output amplitudes in the three ports are given by the input-output relations~\cite{GardinerZoller},
\begin{align}
    a_{\mathrm{out},k} = a_{\mathrm{in},k} - \sqrt{\kappa_k} \alpha \, ,
    \label{eq:inout}
\end{align}
with $k \in \{1,2,3\}$, corresponding to reflection, transmission and absorption respectively, and $a_{\mathrm{in},1} \equiv a_\mathrm{in}$, $a_{\mathrm{in},2} = a_{\mathrm{in},3} = 0$ since we consider the noise to be negligible and take the cavity to be driven only from port 1. In reflection, we then have
\begin{align}
    a_{\mathrm{out},1}^\parallel &= \left(1 - \frac{\kappa_1}{\kappa/2 + i \Delta} \right) a_\mathrm{in}^\parallel \, , \\
    a_{\mathrm{out},1}^\bot &= a_\mathrm{in}^\bot \, ,
\end{align}
for the ampltiudes of the mode-matched and non-mode-matched part, respectively, as the latter effectively never enters the cavity. In general, the output power in any port is given by $P_k = |a_{\mathrm{out},k}|^2$. Specifically, the reflected power is found according to $P_1 = |a_{\mathrm{out},1}^\parallel + a_{\mathrm{out},1}^\bot |^2$. Because the two contributions are by definition orthogonal, $a_\mathrm{in}^\parallel \cdot a_\mathrm{in}^\bot = 0$, it follows
\begin{align}
    P_1 &= |a_{\mathrm{out},1}^\parallel|^2 + |a_{\mathrm{out},1}^\bot|^2 = \epsilon \left[ 1 - \frac{\kappa_1 (\kappa - \kappa_1)}{(\kappa/2)^2 + \Delta^2}\right] P_\mathrm{in} + (1 - \epsilon) P_\mathrm{in} = \nonumber \\ &= \left[1 - \frac{\epsilon \kappa_1(\kappa - \kappa_1)}{(\kappa/2)^2 + \Delta^2} \right] P_\mathrm{in} \, ,
    \label{eq:pow1}
\end{align}
where we have used $P_\mathrm{in} = |a_\mathrm{in}|^2$ for the input power.
The transmitted and absorbed power follow analogously from the corresponding amplitudes, i.e. input-output relations. The situation is simplified compared to reflection in that there is only the (mode-matched) leakage from the cavity. Specifically,
%
\begin{align}
    P_l = \frac{\epsilon \kappa_1 \kappa_l}{(\kappa/2)^2 + \Delta^2} P_\mathrm{in}\, ,
    \label{eq:pow23}
\end{align}
with $l \in \{2,3\}$. The cavity reflection, transmission and absorption coefficients ($\mathcal R$, $\mathcal T$ and $\mathcal A$) are finally given by the ratio of the output power in the pertaining port and the input power. From Eqs~(\ref{eq:pow1}) and (\ref{eq:pow23}), it is straightforward to see their forms are exactly those specified in the main text.

\section{Signal-to-noise ratio}
\label{sec:SI2}

We are interested in ascertaining whether the object is present inside the cavity (state $P$) or not (state $A$). We assume a single detector, placed in either of the two accessible output ports, namely ports 1 and 2. For a general cavity coupling efficiency $\xi$ and a finite object-induced detuning $\Delta_P$, the cavity will reflect and transmit incoming photons in both cases, $A$ and $P$. We thus define our signal in a given port $j$ to be the difference $N_j$ between the photon counts originating from the photons leaking out of the cavity into this port in these two cases, i.e.

\begin{align}
    N_j = |N_A^j - N_P^j| \, ,
\end{align}
where the absolute value is taken simply because the transmission is maximal on resonance (which we take to correspond to state $A$, i.e. $\Delta_A = 0$) and virtually zero in state $P$ assuming $\Delta_P \gg \kappa$, whereas the situation is reversed in reflection. The number of photons leaking out of the cavity in a given port and a given state is related to the input photon number $N_0$ via the corresponding cavity coefficient, but not all of these photons are detected, such that
\begin{align}
    N_X^j = \chi_j N_0 \mathcal J_X \, ,
    \label{eq:goodcounts}
\end{align}
with $\chi_j$ the quantum efficiency of the detector, $X \in \{ A, P\}$ and $\mathcal J = \mathcal R, \mathcal T$ for $j = 1,2$ respectively.

This signal should be measured against the uncertainty in the counts $\Delta N_j$, the total of which contains contributions from both cases with regards to the object's presence, according to
\begin{align}
    \Delta N_j = \sqrt{(\Delta N_A^j)^2 + (\Delta N_P^j)^2} \, .
\end{align}
Assuming the photons obey Poissonian statistics, the uncertainty scales as the square root of the number of counts, with which we have
\begin{align}
    \Delta N_j = \sqrt{N_{\mathrm{tot},A}^j + N_{\mathrm{tot},P}^j} \, ,
\end{align}
where the total counts $N_{\mathrm{tot},X}^j = N_X^j + N_D^j$ are comprised of the ``good'' counts adding to the signal, given by Eq.~(\ref{eq:goodcounts}), as well as the dark counts $N_D^j$. The number of input photons can be expressed as the product $N_0 = C_0 t$ of the input rate $C_0$ and measurement time $t$. Similarly, the number of dark counts is related to the detector dark count rate $C_j$ as $N_D^j = C_j t$.

Putting all of the above together, it follows that the signal-to-noise ratio
\begin{align}
    \mathrm{SNR}_j = \frac{N_j}{\Delta N_j}
\end{align}
is given by Eq.~(2) of the main text,
\begin{align}
 \mathrm{SNR}_j = \frac{ C_0 t \chi_j |{\mathcal J_A} - {\mathcal J_P}| }{\sqrt{C_0t} \cdot \sqrt{\chi_j ({\mathcal J_A} +  {\mathcal J_P})  + 2 C_j/C_0} }
= \frac{ \sqrt{N_0} \chi_j |{\mathcal J_A} - {\mathcal J_P}| }{ \sqrt{\chi_j ({\mathcal J_A} +  {\mathcal J_P})  + 2 D_j} } \, ,
 \end{align}
with $D_j = C_j/C_0$.

\section{Contour plots of $\zeta$ in the $\xi N_0$-space}
\label{sec:SI3}

\begin{figure}[H] 
    \centering
   \includegraphics[scale = \myScale]{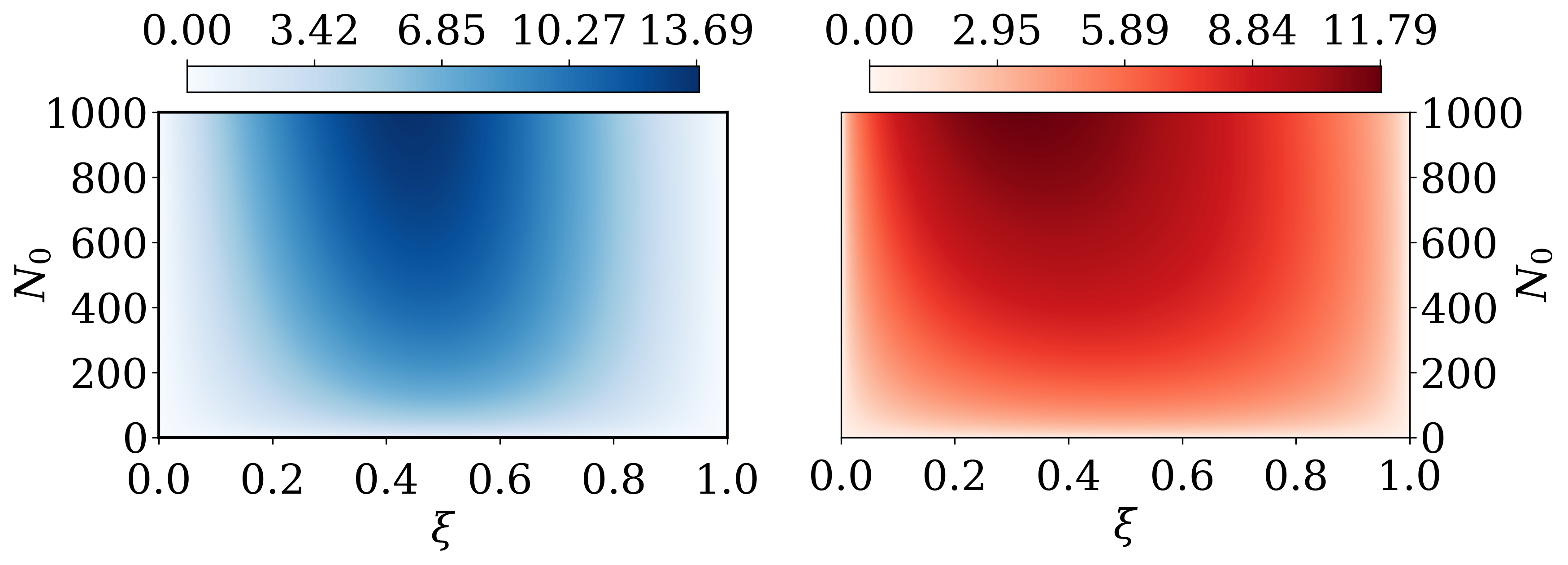}
   \includegraphics[scale = \myScale]{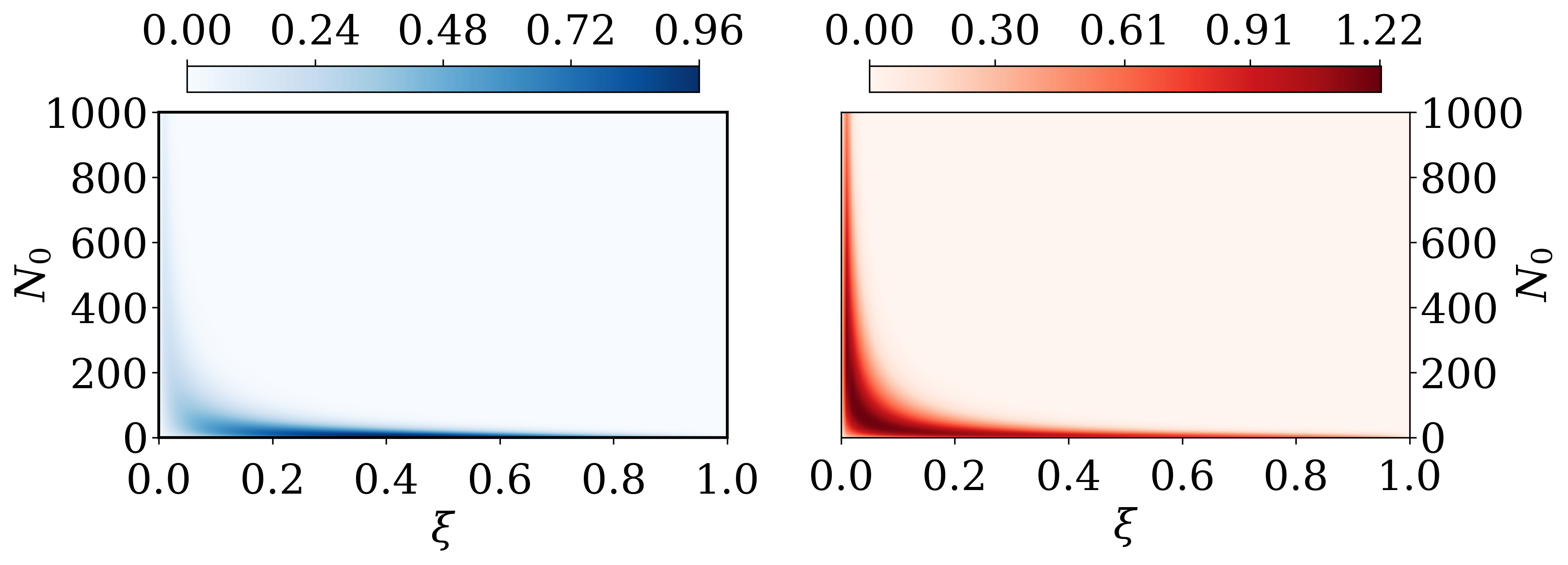}
   \includegraphics[scale = \myScale]{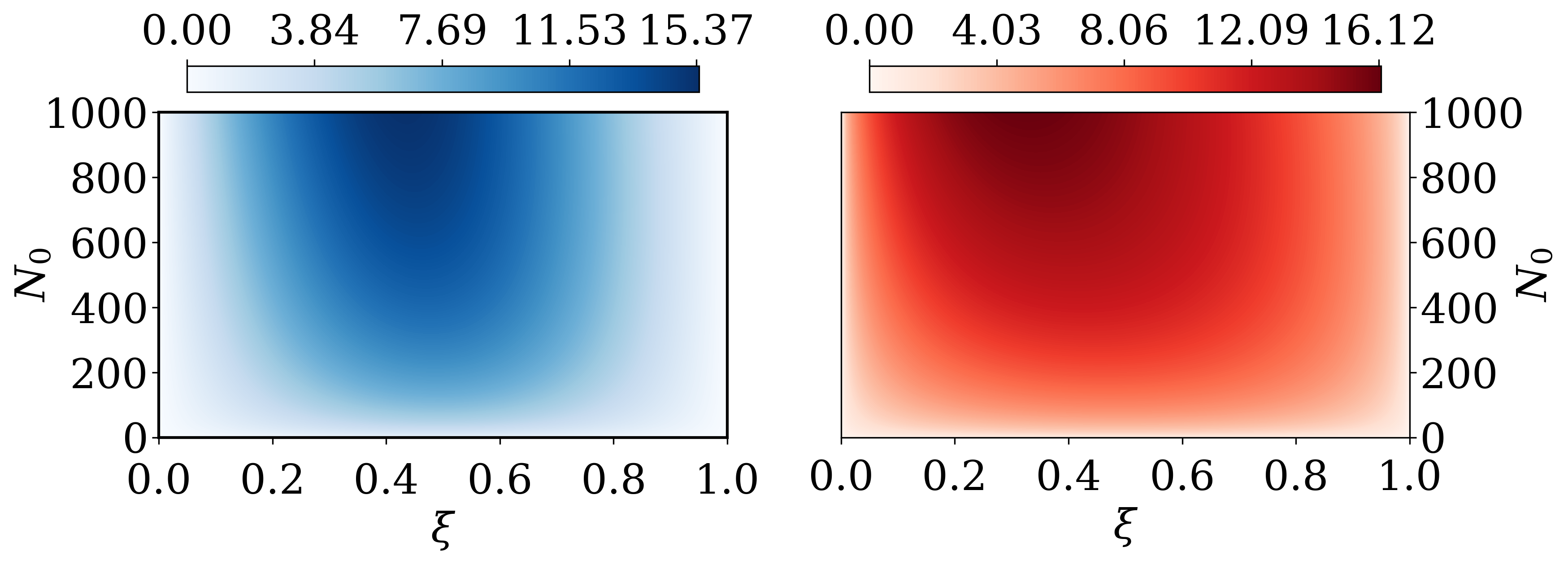}
    \caption{The colorbars represent the products $\zeta_1$ (left) and $\zeta_2$ (right) for $\Delta_P \ll \kappa_A$, i.e. $\kappa_A/(2\pi) = \SI{1.5e7}{\hertz}$, $\Delta_P/(2\pi) = \SI{1.5e4}{\hertz}$, with: 1)~$\kappa_3 \ll \kappa_A$, $\kappa_3=\Delta_P$ (top); 
    2)~$\kappa_3 = \kappa_A$ (middle);
    3)~$\kappa_3 \gg \kappa_A$, $\kappa_3/(2\pi) = \SI{1.5e10}{\hertz}$ (bottom). 
    }
     \label{fig:SI_fig1}
\end{figure}

\begin{figure}[H] 
    \centering
    \includegraphics[scale = \myScale]{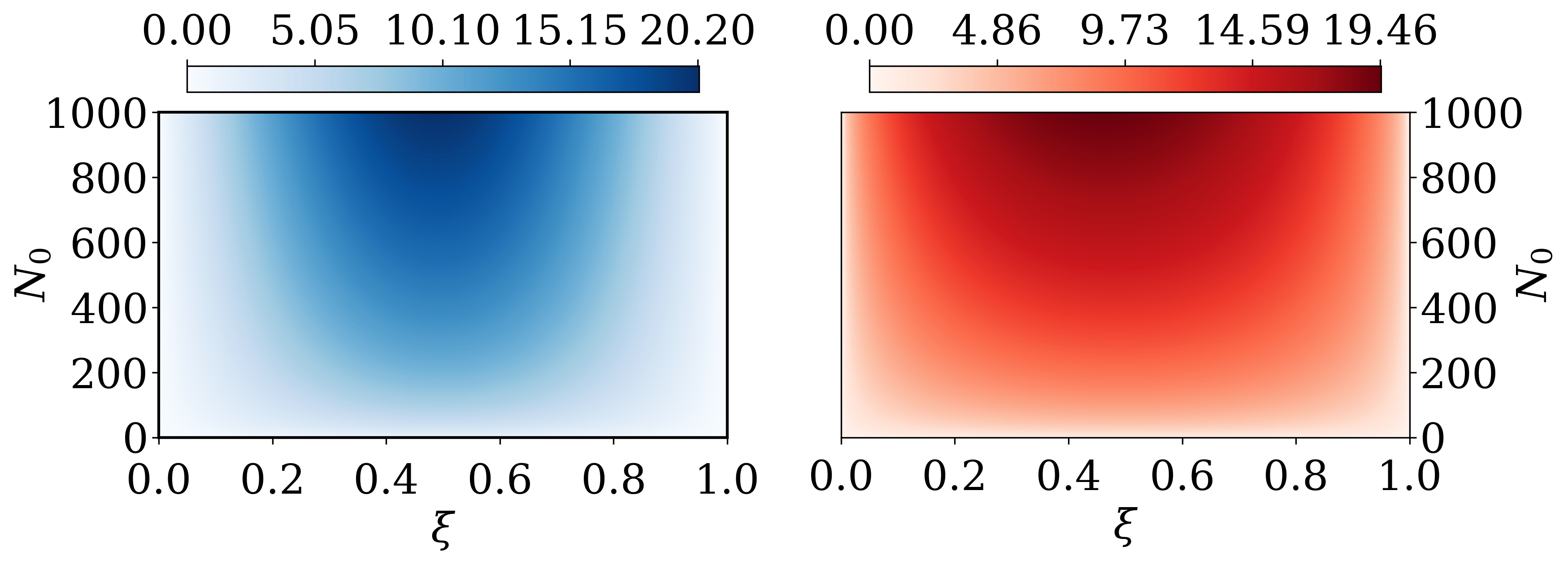}
    \includegraphics[scale = \myScale]{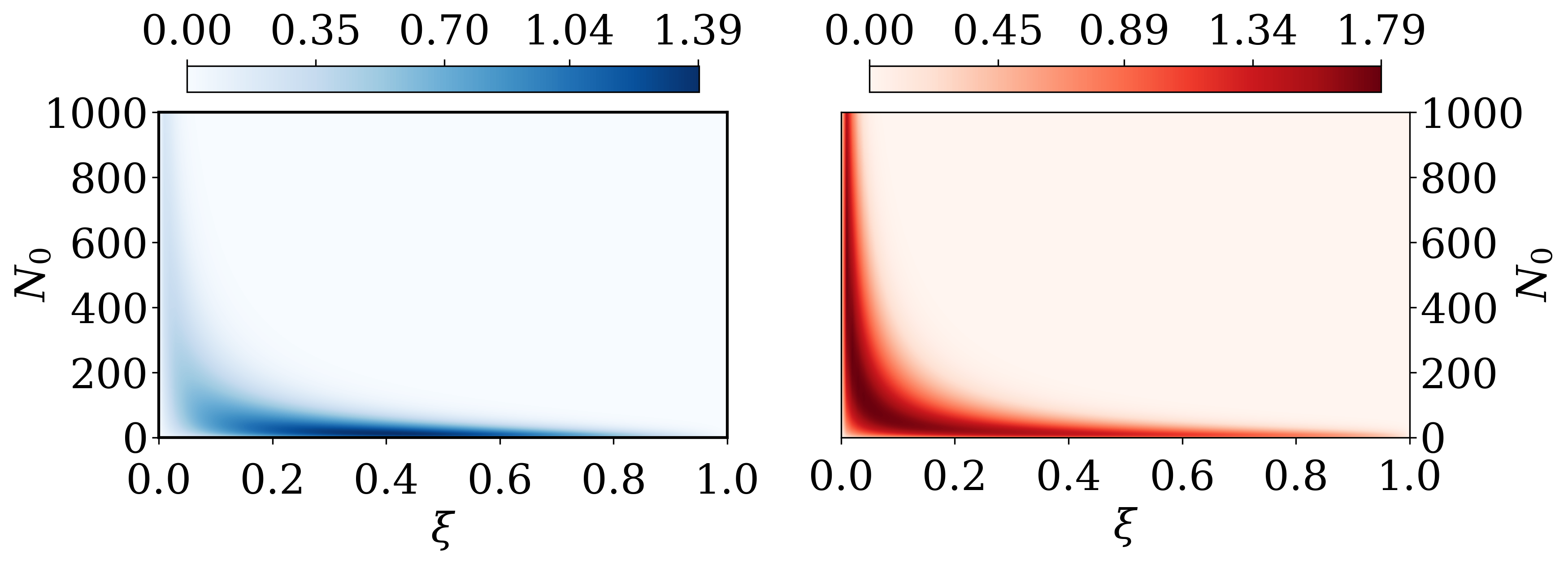}
    \includegraphics[scale = \myScale]{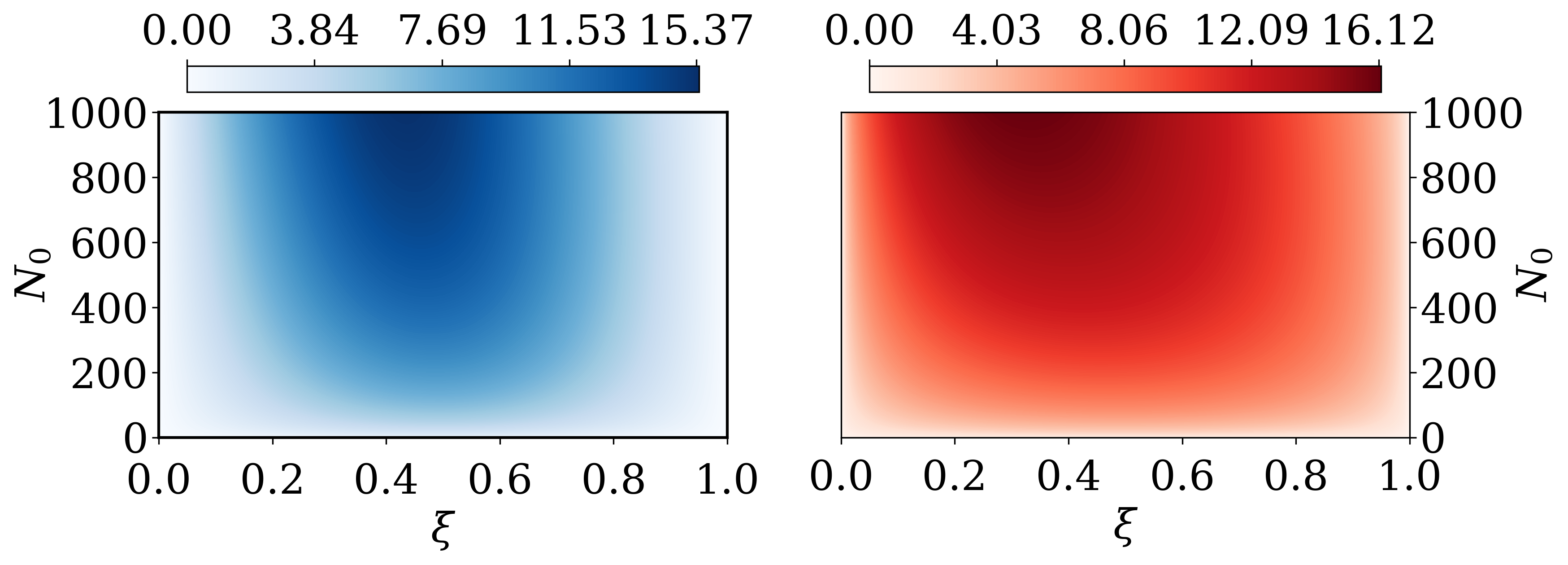}
    \caption{The colorbars represent the products $\zeta_1$ (left) and $\zeta_2$ (right) for $\Delta_P = \kappa_A$, i.e. $\kappa_A/(2\pi) = \Delta_P/(2\pi) = \SI{1.5e7}{\hertz}$, with: 1)~$\kappa_3 \ll \kappa_A$, $\kappa_3/(2\pi) = \SI{1.5e4}{\hertz}$ (top); 
    2)~$\kappa_3 = \kappa_A = \Delta_P$ (middle);
    3)~$\kappa_3 \gg \kappa_A$, $\kappa_3/(2\pi) = \SI{1.5e10}{\hertz}$ (bottom). 
    }
    \label{fig:SI_fig2}
\end{figure}

\begin{figure}[H] 
    \centering
    \includegraphics[scale = \myScale]{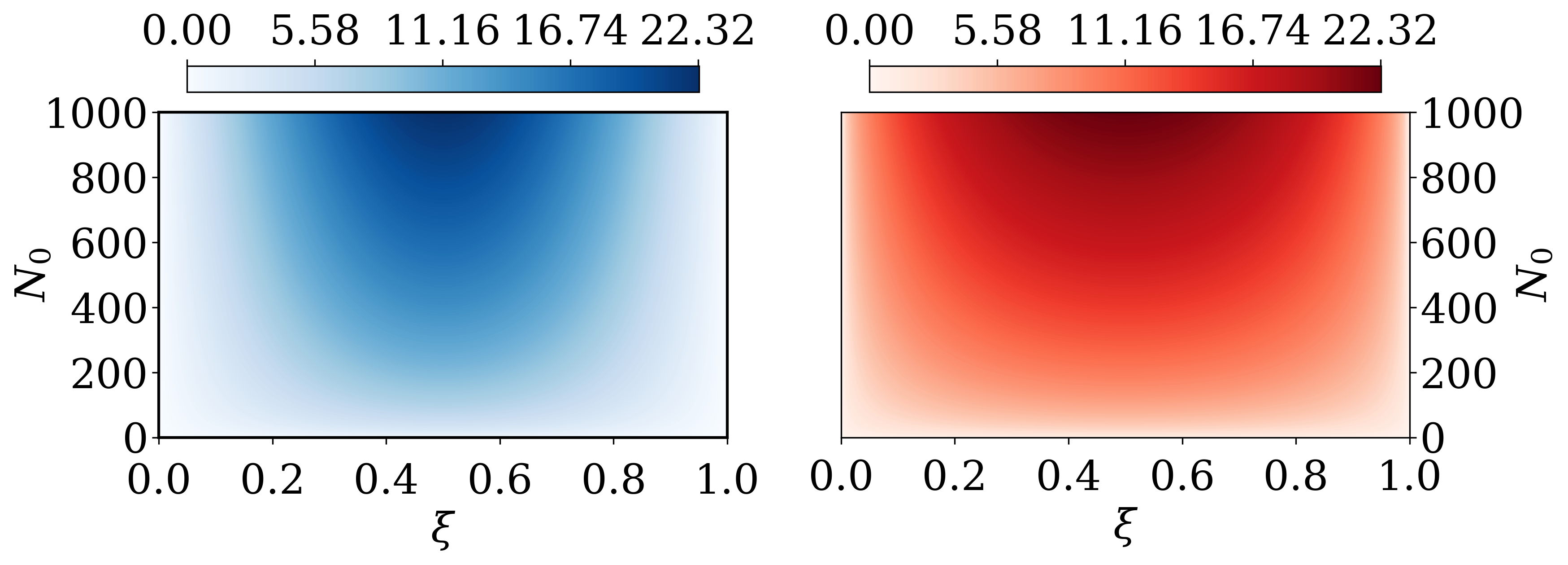}
    \includegraphics[scale = \myScale]{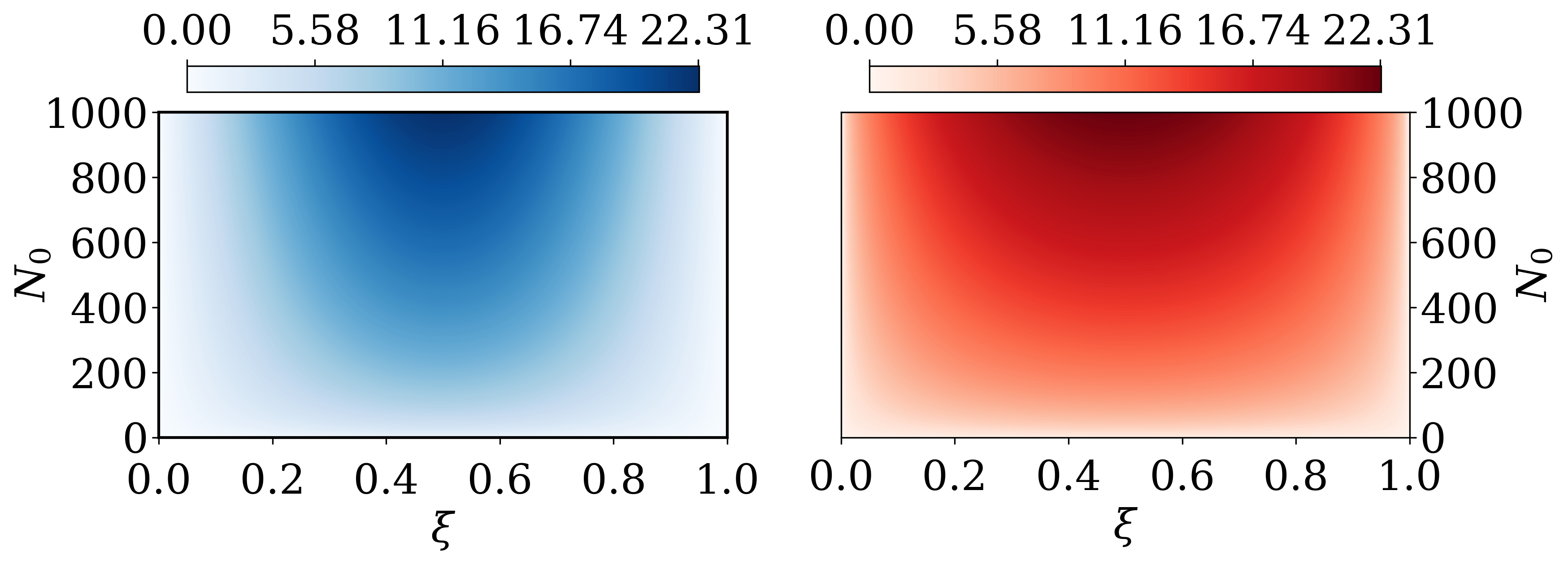}
    \includegraphics[scale = \myScale]{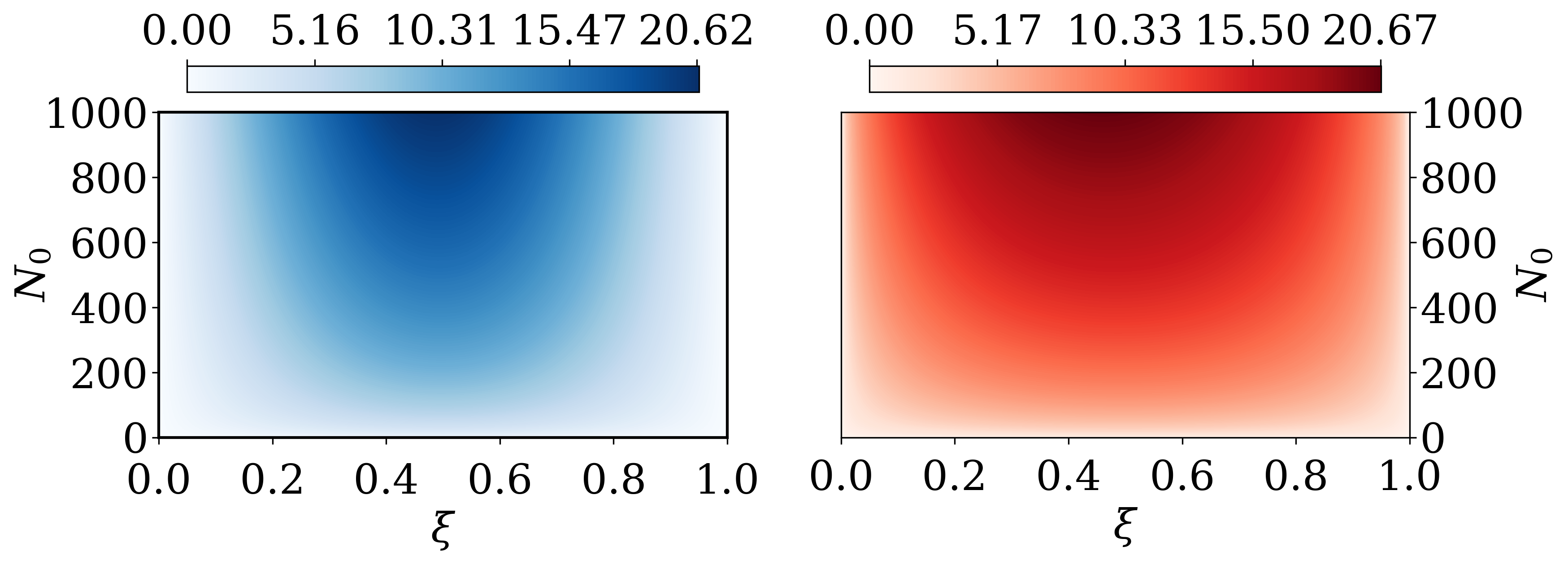}
    \caption{The colorbars represent the products $\zeta_1$ (left) and $\zeta_2$ (right) for $\Delta_P \gg \kappa_A$, i.e. $\kappa_A/(2\pi) = \SI{1.5e7}{\hertz}$, $\Delta_P/(2\pi) = \SI{1.5e10}{\hertz}$, with: 1)~$\kappa_3 \ll \kappa_A$, $\kappa_3/(2\pi) = \SI{1.5e4}{\hertz}$ (top); 
    2)~$\kappa_3 = \kappa_A$ (middle);
    3)~$\kappa_3 \gg \kappa_A$, $\kappa_3 = \Delta_P$ (bottom). 
    }
     \label{fig:SI_fig3}
\end{figure}

\vspace{-20mm}
Here we discuss the plots of the products of security and signal-to-noise ratio ($\zeta_1$ and $\zeta_2$), analogous to those Fig.~\ref{fig:etaS_SNRs_independed-2Cases}(c) and \ref{fig:etaS_SNRs_independed-2Cases}(d) of the main text but for a range of the system parameters' values, in order to study the concomitant changes in the positions of the products' global maxima in $\xi N_0$-space. Specifically, we fix $\kappa_A/(2\pi) = \SI{1.5e7}{\hertz}$ and $\epsilon_P=0.2$, and vary $\Delta_P$ and $\kappa_3$ in a ``nested'' manner: for each of the regimes $\Delta_P \ll \kappa_A $, $\Delta_P \approx \kappa_A$ and $\Delta_P \gg \kappa_A$, shown in Figs.~\ref{fig:SI_fig1}--\ref{fig:SI_fig3} respectively, we vary $\kappa_3$ as $\kappa_3 \ll \kappa_A$, $ \kappa_3 \approx \kappa_A$ and $\kappa_3 \gg \kappa_A$. In these figures, the three regimes for either variable are represented by the values $\left(\Delta_P/(2\pi), \kappa_3/(2\pi)\right)\in \{ \numlist[list-final-separator = {, }]{1.5e4;1.5e7;1.5e10} \} \, \si{\hertz}$. The detector efficiencies are fixed to $\chi_1 = \chi_2 = 0.5$ and their dark count ratios with respect to the input flux to $D_1 = D_2 = \num{e-3}$, as in the main text.

\begin{figure}[H]
    \centering
    \includegraphics[scale = \myScaleSmall]{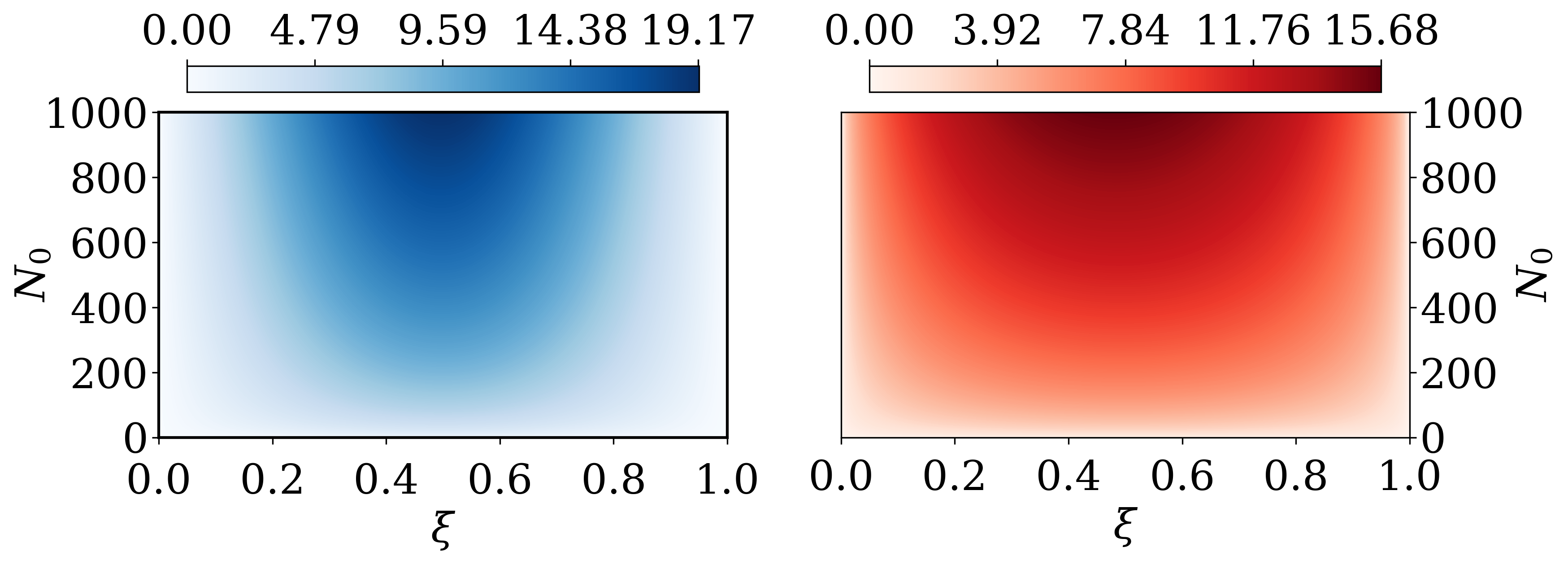}
    \includegraphics[scale = \myScaleSmall]{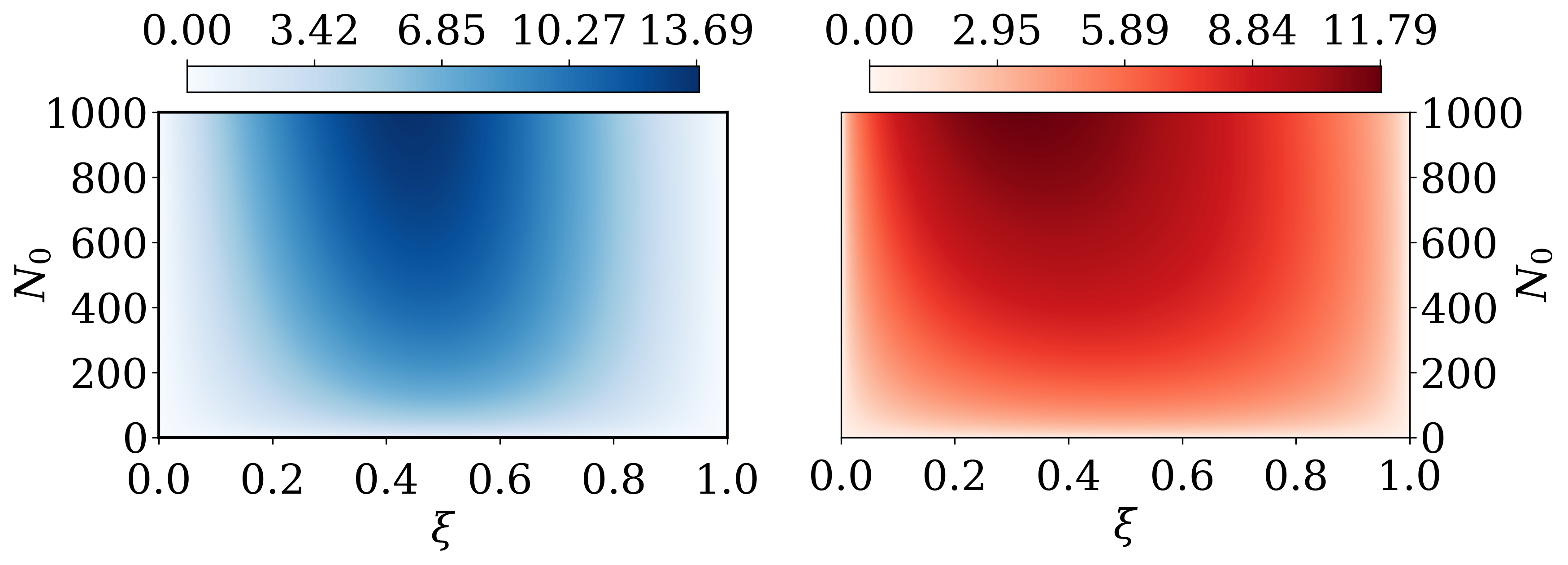}
    \includegraphics[scale = \myScaleSmall]{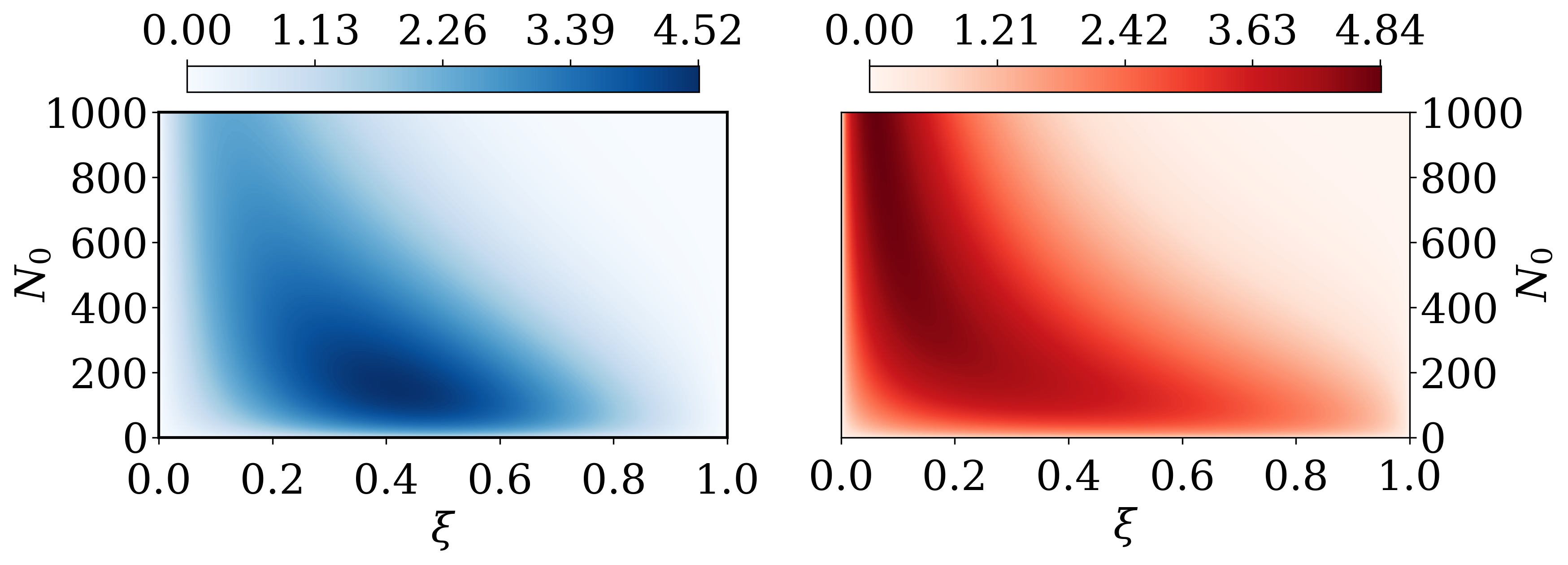}
    \includegraphics[scale = \myScaleSmall]{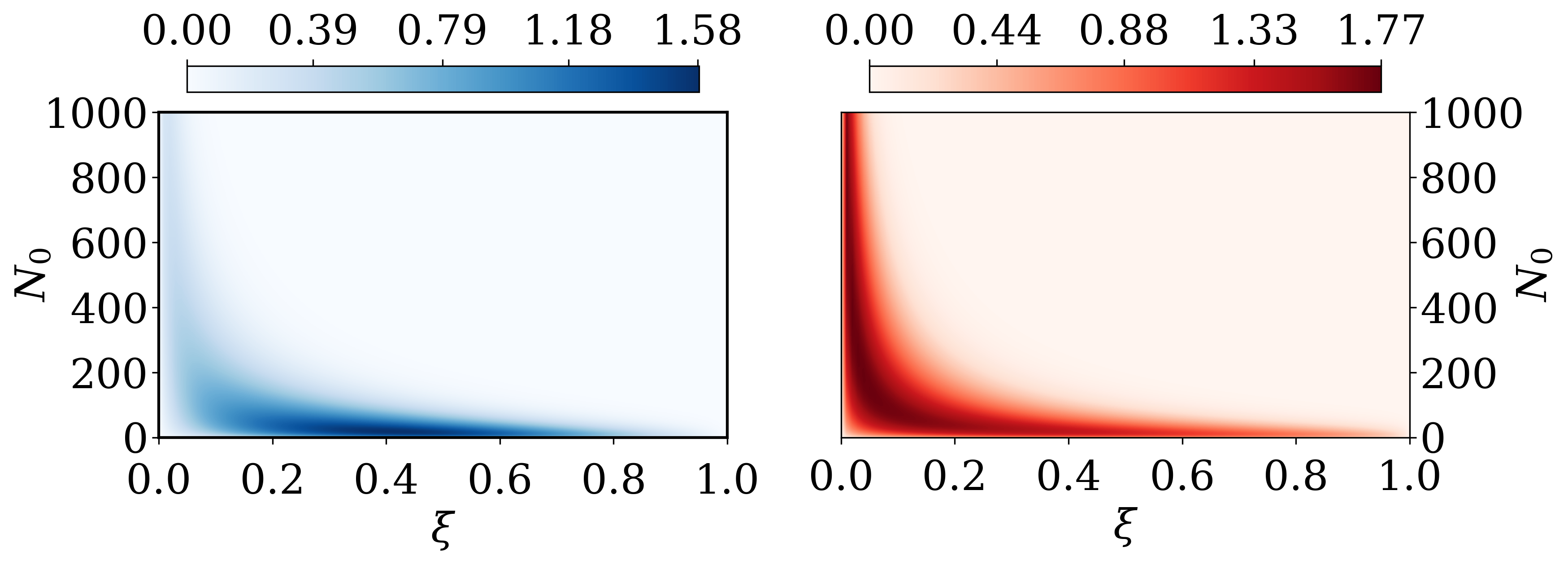}
    \includegraphics[scale = \myScaleSmall]{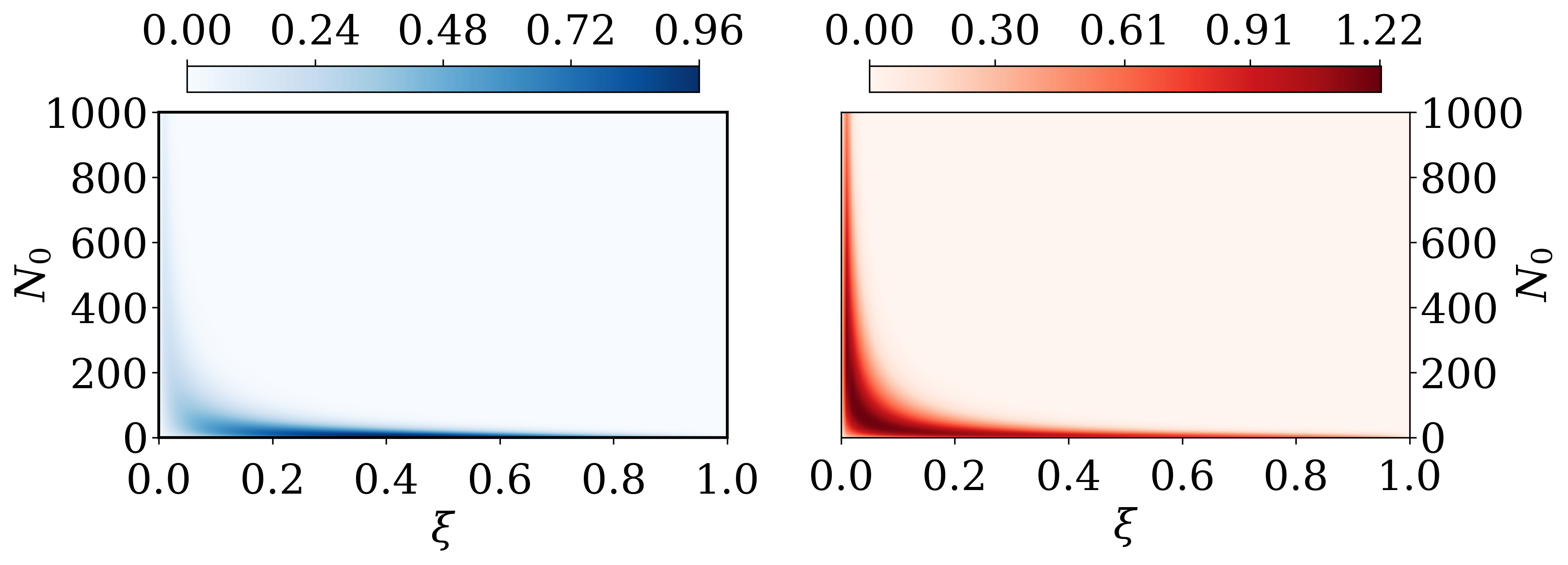}
    \includegraphics[scale = \myScaleSmall]{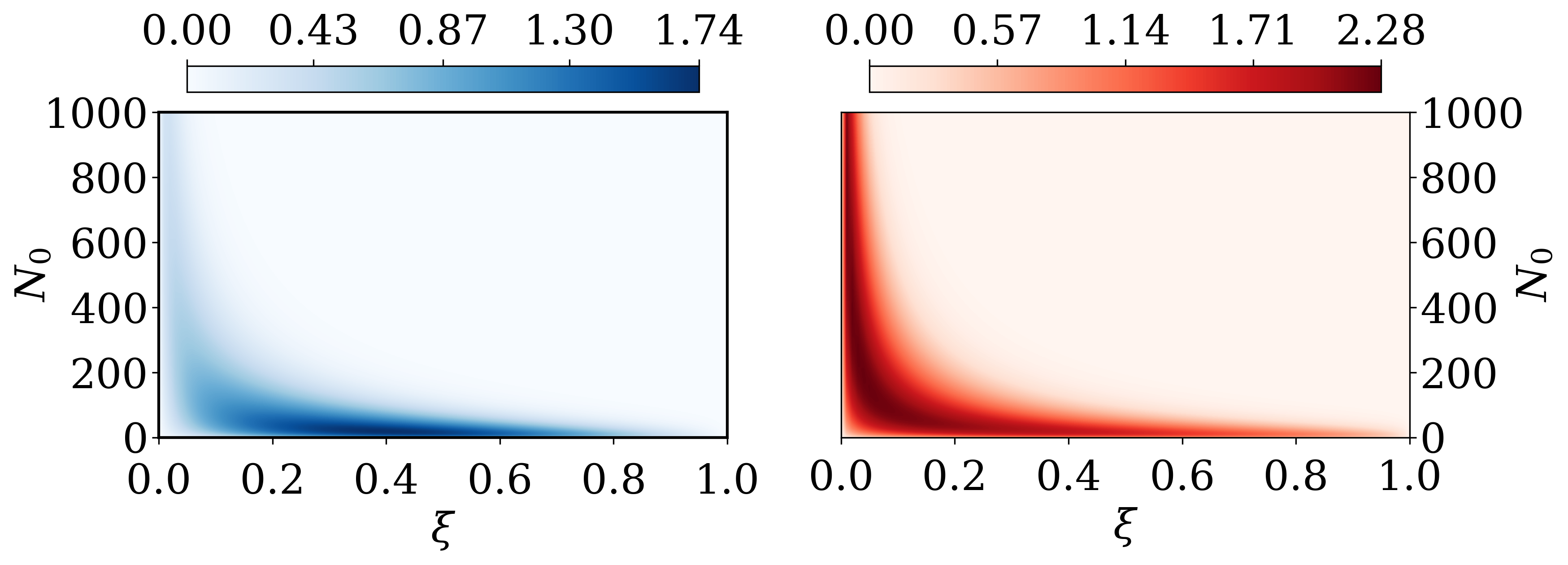}
    \includegraphics[scale = \myScaleSmall]{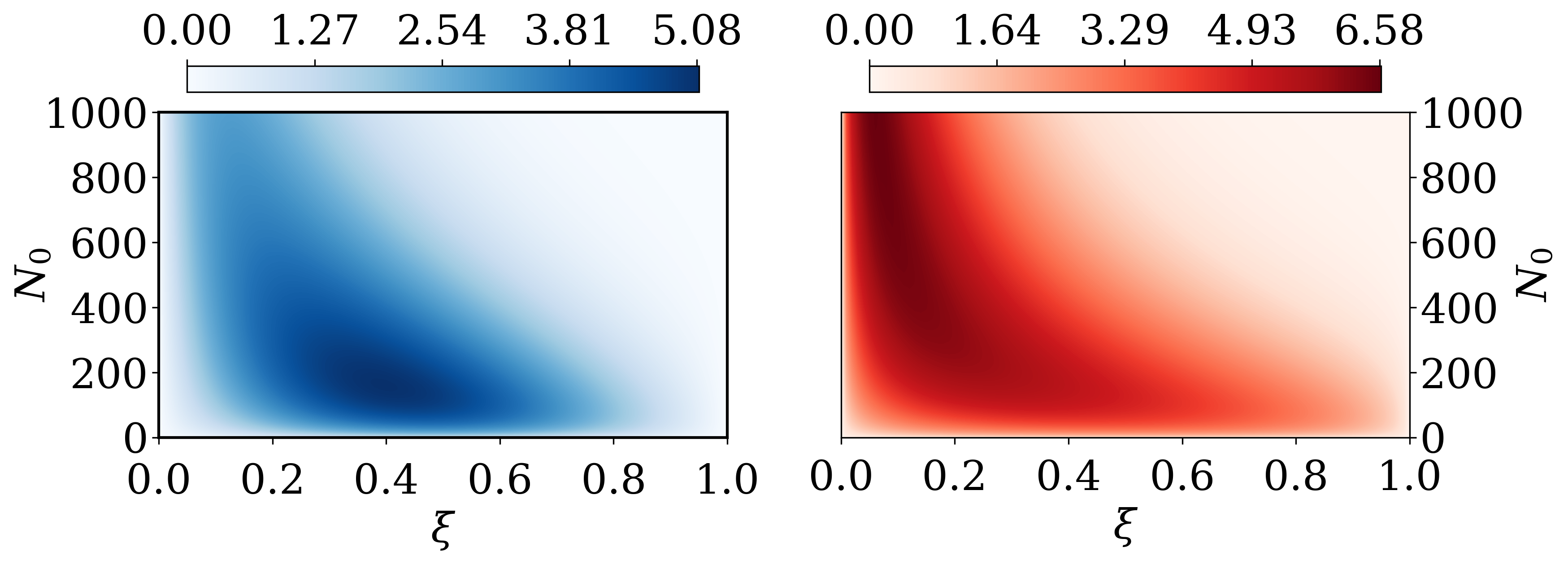}
    \includegraphics[scale = \myScaleSmall]{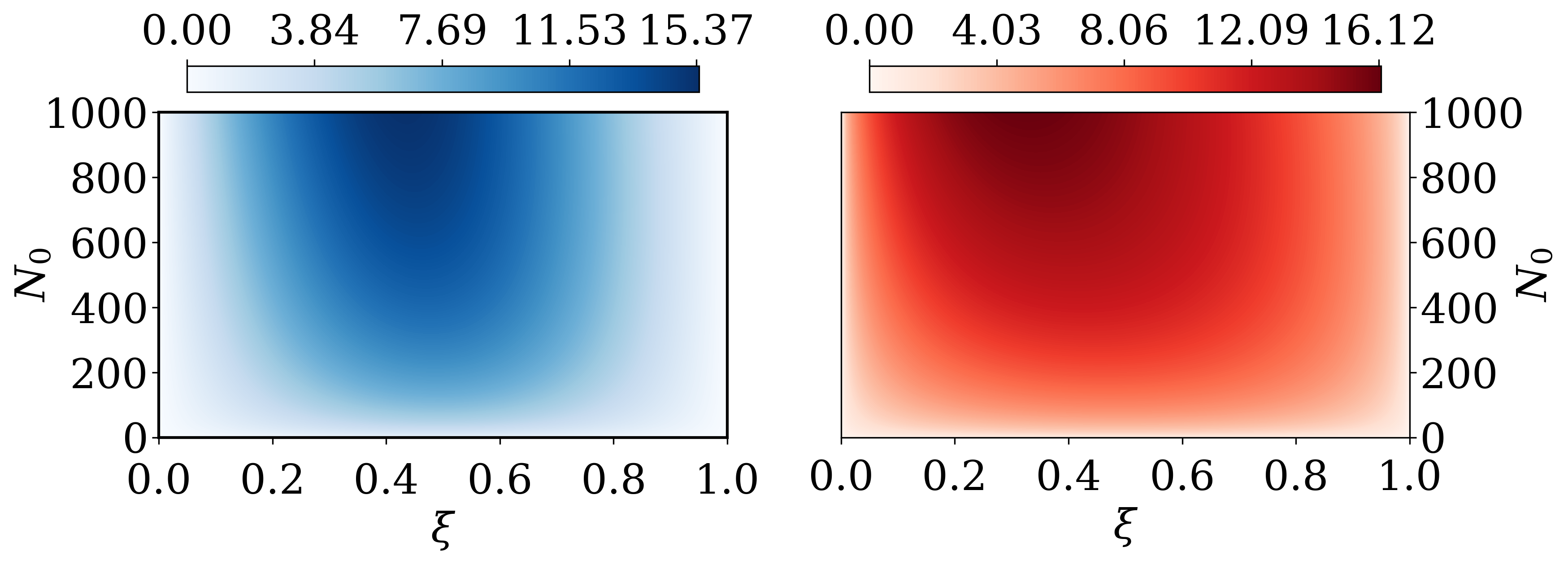}
    \caption{The products $\zeta_1$ (blue colorbar) and $\zeta_2$ (red colorbar) for $\Delta_P \ll \kappa_A$, specifically with $\kappa_A/(2\pi) = \SI{1.5e7}{\hertz}$ and $\Delta_P/(2\pi) = \SI{1.5e4}{\hertz}$. The absorption loss rate is varied in the range $\kappa_3/(2\pi) \in [\num{1.5e3}, \num{1.5e10}] \, \si{\hertz}$, increasing in steps of one decade, first from left to right within each row (while kept the same for a given blue-red pair) and then row-wise from top to bottom.
    }
    \label{fig:SI_fig4}
\end{figure}

\begin{figure}[H]
    \centering
    \includegraphics[scale = \myScaleSmall]{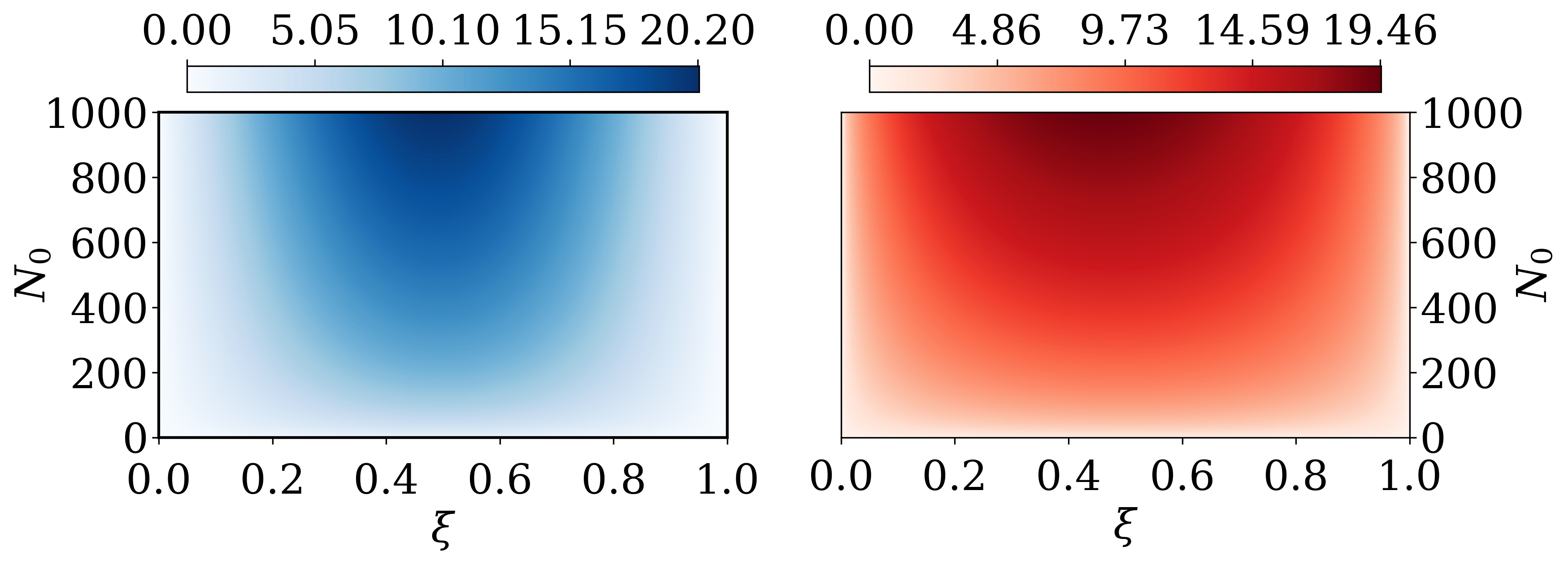}
    \includegraphics[scale = \myScaleSmall]{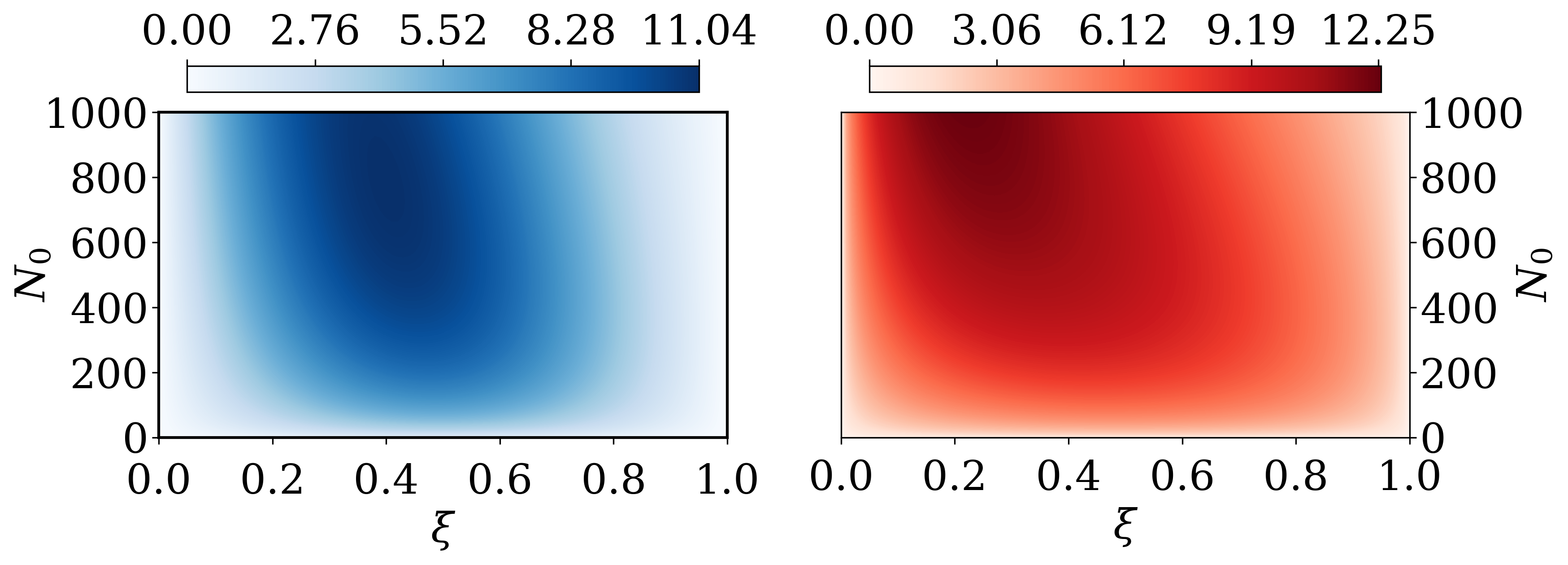}
    \includegraphics[scale = \myScaleSmall]{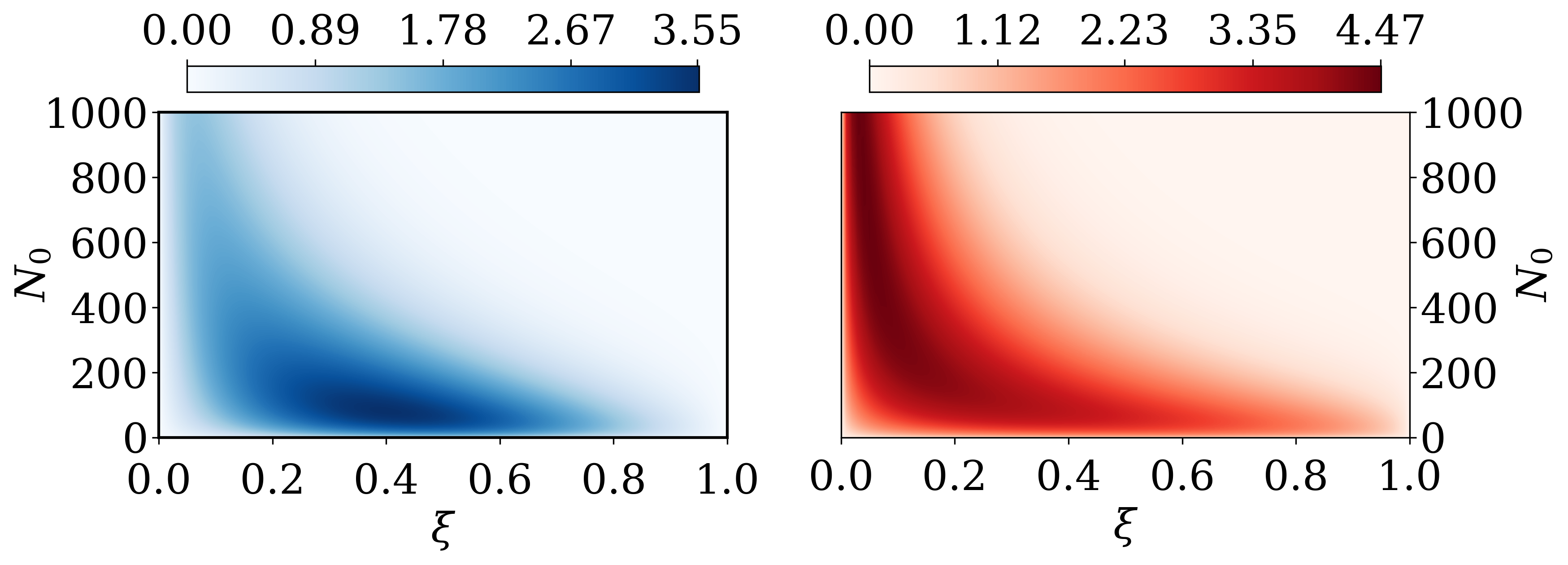}
    \includegraphics[scale = \myScaleSmall]{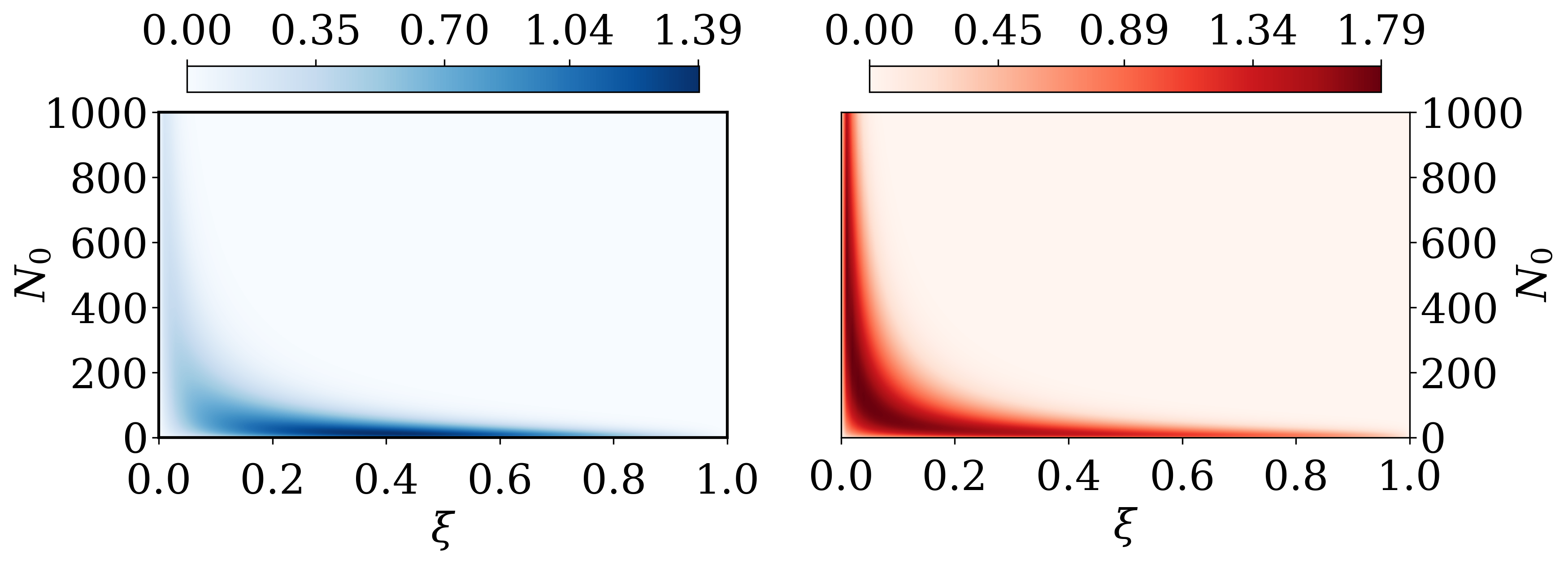}
    \includegraphics[scale = \myScaleSmall]{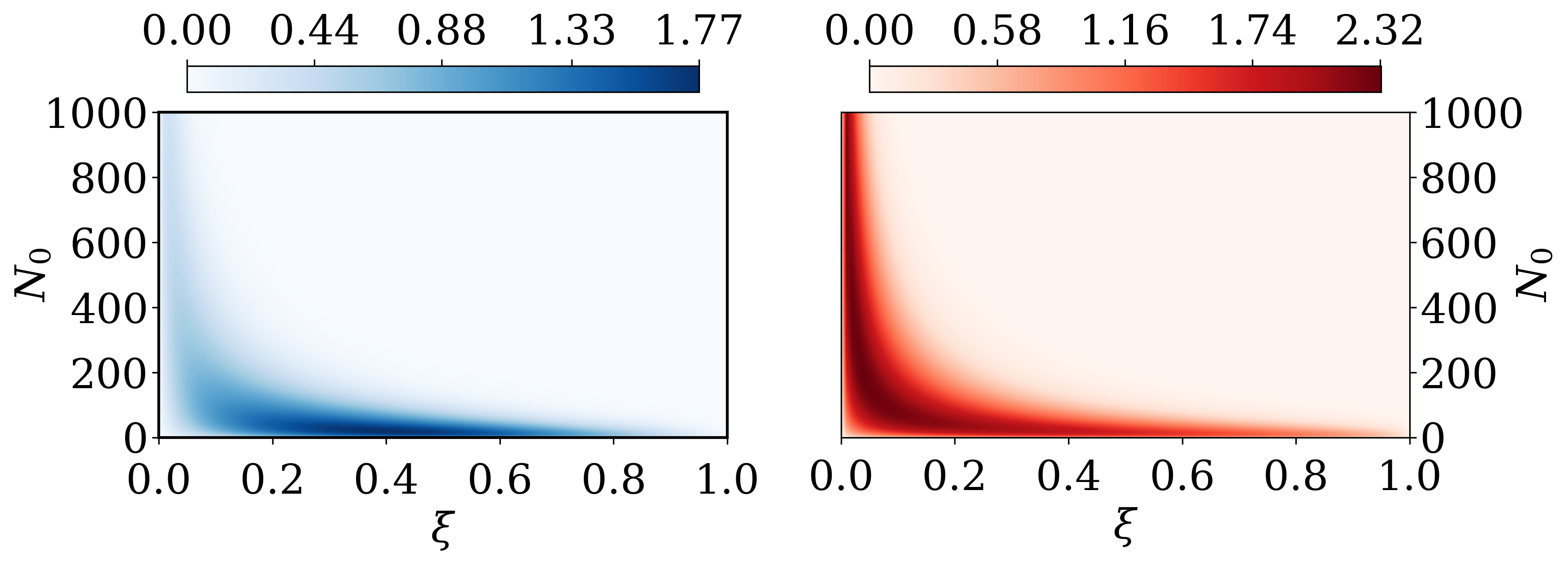}
    \includegraphics[scale = \myScaleSmall]{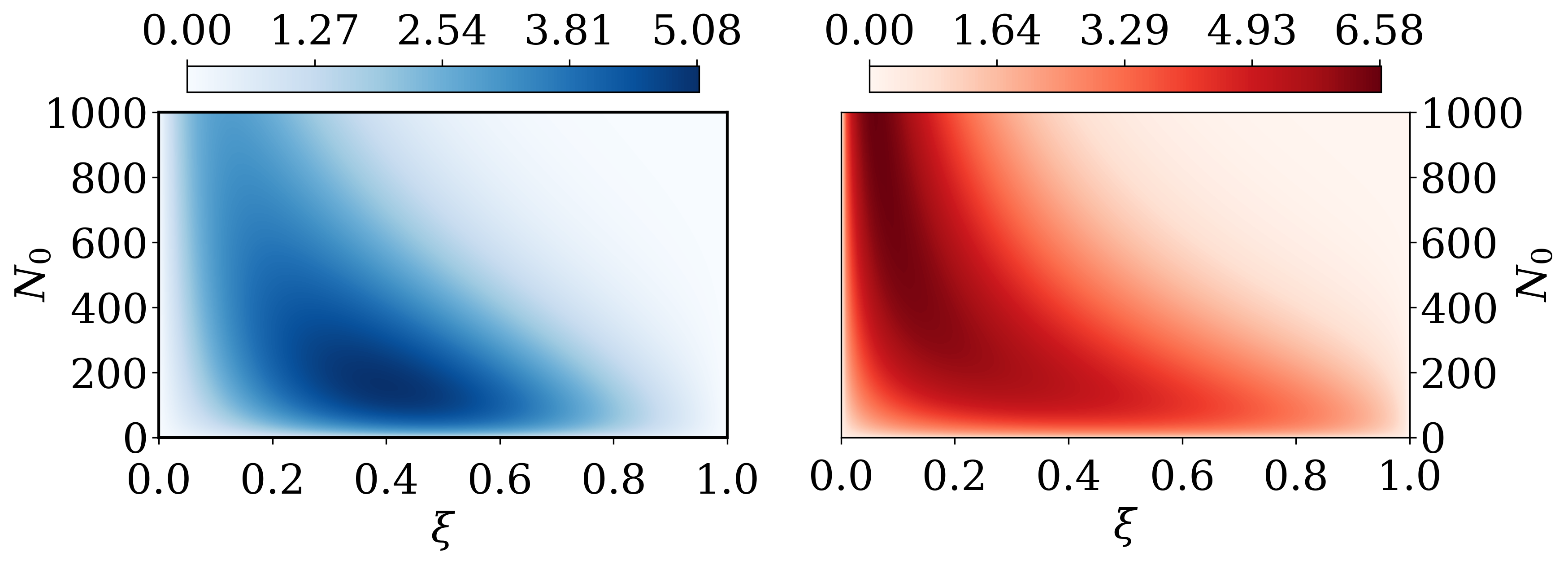}
    \includegraphics[scale = \myScaleSmall]{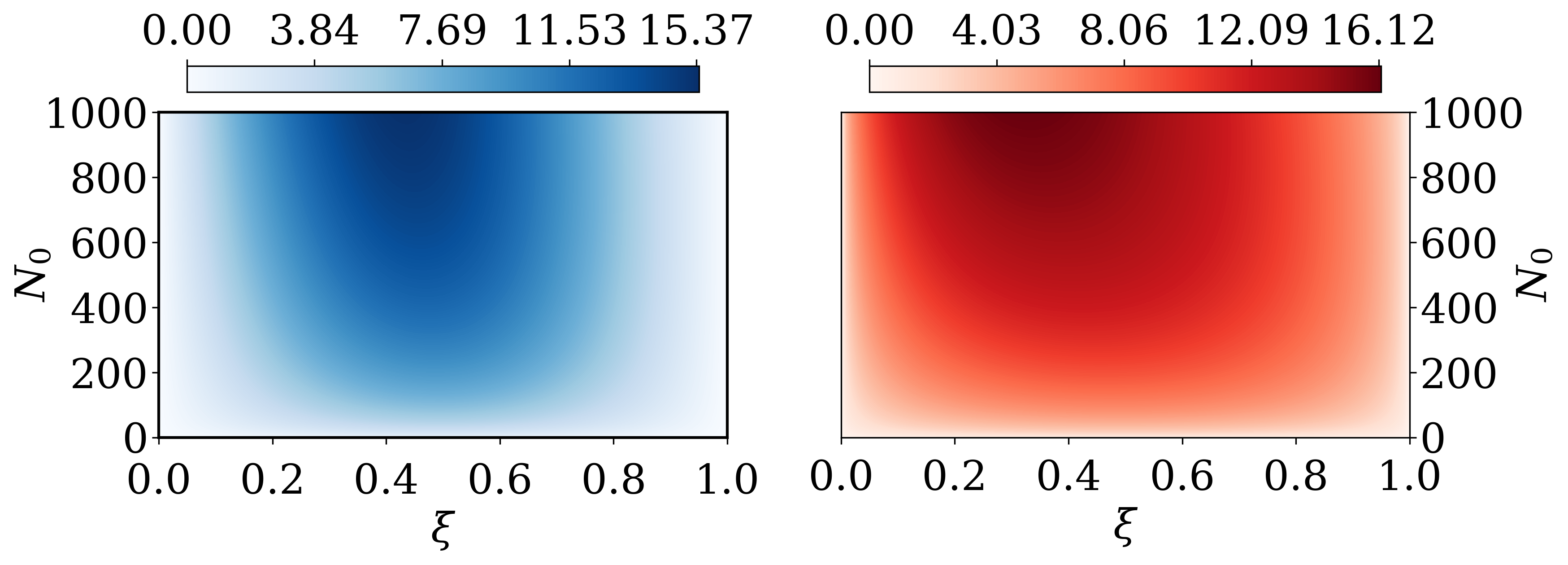}
    \includegraphics[scale = \myScaleSmall]{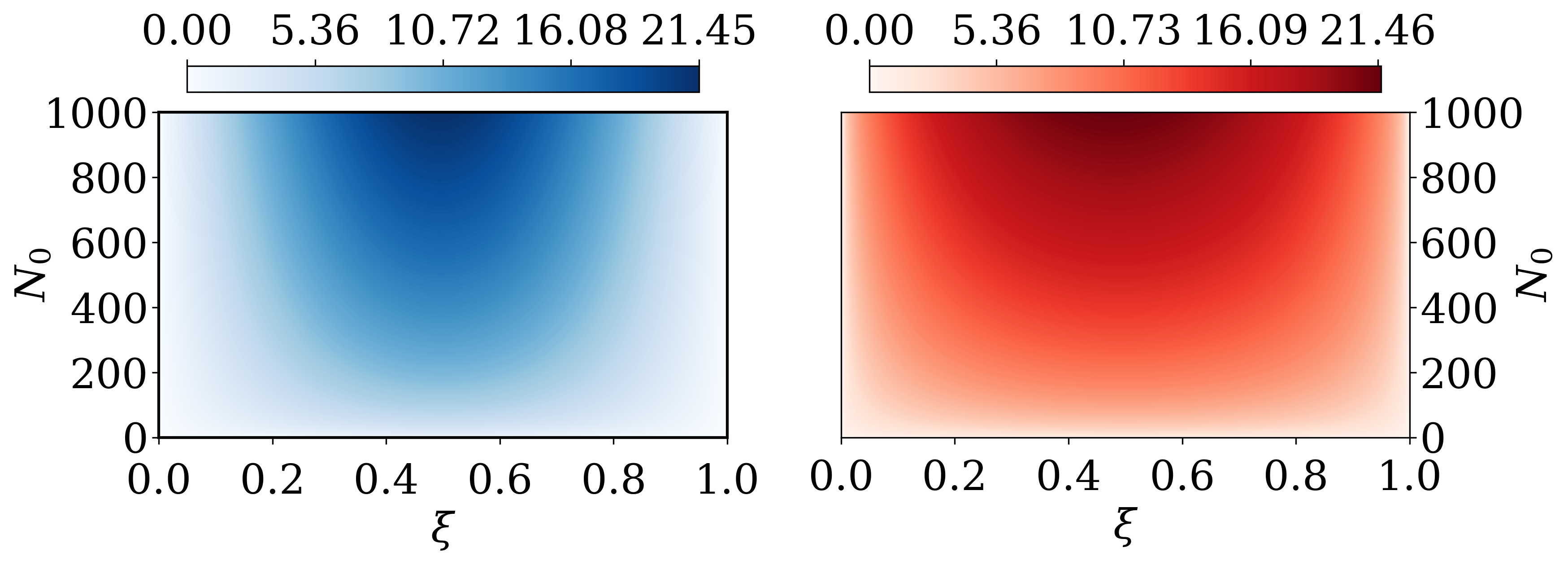}
    \caption{The products $\zeta_1$ (blue colorbar) and $\zeta_2$ (red colorbar) for $\Delta_P/(2\pi) = \kappa_A/(2\pi) = \SI{1.5e7}{\hertz}$. The absorption loss rate is varied in the range $\kappa_3/(2\pi) \in [\num{1.5e4}, \num{1.5e11}] \, \si{\hertz}$, increasing in steps of one decade, first from left to right within each row (while kept the same for a given blue-red pair) and then row-wise from top to bottom.
    }
    \label{fig:SI_fig5}
\end{figure}

The middle rows of Figs.~\ref{fig:SI_fig1} and \ref{fig:SI_fig2} clearly pertain to parameter regimes where the maxima of both $\zeta_1$ and $\zeta_2$ are shifted in $\xi$ towards an undercoupled cavity (and towards a lower photon number $N_0$). The shift in $\xi$ is summarized in Figs.~\ref{fig:etaS_SNRs_product_surfacePlot}(c) and \ref{fig:etaS_SNRs_product_surfacePlot}(d) of the main text, where it is indeed apparent around $\kappa_3 \sim \kappa_A$ for $\Delta_P \lesssim \kappa_A$, but not for $\Delta_P \gtrsim \kappa_A$ (the extreme case of which corresponds to Fig.~\ref{fig:SI_fig3}). In order to study this more carefully, we now focus on the cases $\Delta_P \ll \kappa_A$ and $\Delta_P = \kappa_A$ and vary $\kappa_3$ in finer steps compared to those above. The results are shown in Figs.~\ref{fig:SI_fig4} and \ref{fig:SI_fig5} respectively, with $\kappa_A$ and the detector specifications fixed the same as before. There are now intermediate shapes which confirm that the contour plots undergo smooth transitions from the typically expected semi-oval, to a distorted boomerang-like shape, and back. For extreme values of $\kappa_3$, both $\zeta_1$ and $\zeta_2$ reach their maxima close to 
$\xi = 0.5$ and $N_0 = 1000$, in both figures. The maximal distortion of the shape occurs around $\kappa_3 \approx \kappa_A$, with the pull towards an undercoupled cavity significantly more pronounced in the case of transmission~($\zeta_2$), where the maximum appears at $\xi \approx 0.03$. Reflection~($\zeta_1$) is instead more flattened in $N_0$, with the coupling efficiency deviating (only) up to $\xi \approx 0.4$, but the optimal photon number going all the way down to a few tens, compared to a few hundred for transmission.

\newpage

\end{document}